\pdfoutput=1

\documentclass[a4paper,11pt]{article}

\usepackage[affil-sl,auth-lg]{authblk}
\usepackage{amssymb,amsmath,amsbsy}
\usepackage{mathrsfs}
\usepackage{mathtools}

\usepackage{dsfont}
\usepackage{appendix}
\usepackage[usenames, dvipsnames]{color}

\usepackage{parskip}

\usepackage{xspace}
\usepackage{braket}
\usepackage{array}
\usepackage{feynmp}
\usepackage{pinlabel}
\usepackage[top=1.2in, bottom=1.2in,left=0.8in, right=0.8in,includefoot]{geometry}

\definecolor{nicered}{rgb}{0.7,0.1,0.1}
\definecolor{nicegreen}{rgb}{0.1,0.5,0.1}

\definecolor{rosso}{cmyk}{0,1,1,0.4}
\definecolor{babypink}{rgb}{0.96, 0.76, 0.76}
\definecolor{babyblueeyes}{rgb}{0.63, 0.79, 0.95}
\definecolor{azure(colorwheel)}{rgb}{0.0, 0.5, 1.0}
\definecolor{amethyst}{rgb}{0.6, 0.4, 0.8}

\usepackage[backref=none]{hyperref}
\definecolor{MyDarkBlue}{rgb}{0,0.1,0.7}
\hypersetup{pdfborder={0 0 0},colorlinks,breaklinks=true,
  urlcolor={MyDarkBlue},citecolor={nicegreen},linkcolor={amethyst}
}

\usepackage{graphicx,rotating,colortbl}

\usepackage{verbatim}
\usepackage{doi}
\usepackage{url}
\usepackage{array}

\usepackage{cite}
\usepackage{float}
\usepackage[labelfont=bf]{caption}
\usepackage{subcaption}
\usepackage{parskip}
\usepackage{hhline}

\usepackage{titlesec}
\titlelabel{\thetitle.\quad}

\def\Eq#1{Eq.~(\ref{#1})}

\newcommand{\newc}{\newcommand}
\newc{\be}{\begin{equation}}
\newc{\ee}{\end{equation}}
\newc{\bal}{\begin{align}}
\newc{\eal}{\end{align}}
\newc{\ba}{\begin{eqnarray}}
\newc{\ea}{\end{eqnarray}}
\newc{\bea}{\begin{eqnarray*}}
\newc{\eea}{\end{eqnarray*}}
\newc{\D}{\partial}
\newc{\shh}{\lambda_{h}}
\newc{\shf}{\lambda_{h\phi}}
\newc{\sff}{\lambda_{\phi}}
\newc{\sss}{\lambda_{s}}
\newc{\ssisi}{\lambda_{\sigma}}
\newc{\sssi}{\lambda_{s \sigma}}
\newc{\shs}{\lambda_{h s}}
\newc{\shsi}{\lambda_{h \sigma}}
\newc{\sfs}{\lambda_{\phi s}}
\newc{\sfsi}{\lambda_{\phi \sigma}}
\newc{\shz}{\lambda_{h}^{(0)}}
\newc{\shfz}{\lambda_{h\phi}^{(0)}}
\newc{\sfiz}{\lambda_{\phi}^{(0)}}
\newc{\sssz}{\lambda_{s}^{(0)}}
\newc{\ssfz}{\lambda_{s\phi}^{(0)}}
\newc{\som}{\sin\omega}
\newc{\com}{\cos\omega}
\newc{\sth}{\sin\theta}
\newc{\cth}{\cos\theta}
\newc{\stom}{\sin^2\omega}
\newc{\ctom}{\cos^2\omega}
\newc{\stth}{\sin^2\theta}
\newc{\ctth}{\cos^2\theta}
\newc{\ie}{{\it i.e.} }
\newc{\eg}{{\it e.g.} }
\newc{\etc}{{\it etc.} }
\newc{\etal}{{\it et al.}}

\newcommand{\eV}{\; \mathrm{eV}}

\newcommand{\GeV}{\; \mathrm{GeV}}
\newcommand{\TeV}{\; \mathrm{TeV}}
\newcommand{\lapproxeq}{\lower .7ex\hbox{$\;\stackrel{\textstyle
<}{\sim}\;$}}
\newcommand{\gapproxeq}{\lower .7ex\hbox{$\;\stackrel{\textstyle
>}{\sim}\;$}}
\newcommand{\stackdown}[2]{\lower 1.4ex\hbox{$\;\stackrel{\textstyle{#1}}
{\scriptstyle{#2}}\;$}}

\numberwithin{equation}{section}

\begin{document}

\title{\textbf{Dark matter and neutrino masses from a scale-invariant multi-Higgs portal}}

\author{{A}lexandros {K}aram\footnote{email: {\href{mailto:alkaram@cc.uoi.gr}{alkaram@cc.uoi.gr}}}~  and 
Kyriakos Tamvakis\footnote{email: {\href{mailto:tamvakis@uoi.gr}{tamvakis@uoi.gr}}}}

\affil{Department of Physics, Division of Theoretical Physics, 
 University of Ioannina,  GR 45110 Ioannina, Greece}

\maketitle

\begin{abstract}
We consider a classically scale invariant version of the Standard Model, extended by an extra dark $SU(2)_X$ gauge group. Apart from the dark gauge bosons and a dark scalar doublet which is coupled to the Standard Model Higgs through a portal coupling, we incorporate right-handed neutrinos and an additional real singlet scalar field. After symmetry breaking \`{a} la Coleman-Weinberg, we examine the multi-Higgs sector and impose theoretical and experimental constraints. In addition, by computing the dark matter relic abundance and the spin-independent scattering cross section off a nucleon we determine the viable dark matter mass range in accordance with present limits. The model can be tested in the near future by collider experiments and direct detection searches such as XENON 1T.
\end{abstract}

\newpage
\section{Introduction}
\label{intro}
In 2012, the ATLAS and CMS experiments of the Large Hadron Collider (LHC) at CERN reported the discovery~\cite{Aad2012,Chatrchyan2012} of a boson that appears to be~\cite{Aad2013,Chatrchyan2013} the long-sought Higgs particle~\cite{Englert1964,Higgs1964,Higgs1964a,Guralnik1964} of the Standard Model (SM)~\cite{Glashow1961,Weinberg1967,Salam1968}. The latest precise measurements of the Higgs boson mass \cite{Aad2015} place its value at $M_h = 125.09 \pm 0.24 \,\,\GeV$ . With the last piece of the puzzle now in place the SM seems complete. 

Although the SM is a perfectly consistent quantum field theory and is expected to be valid up to energies of the order of the Planck scale $(M_P)$, where quantum gravitational corrections are believed to come into play, with the observed value of the Higgs boson mass the Higgs quartic self-coupling runs to negative values at an energy scale lower than $M_P$. This is the SM \textit{vacuum stability problem}. Furthermore, if one considers the SM as an effective field theory embedded in a more fundamental theory involving other scales, then the issue of the smallness of the Higgs mass in comparison to $M_P$ or other scales (like $M_{\text{GUT}}$) arises and, as a result, the so-called {\textit{hierarchy problem}}. In a general UV completion of the SM the Higgs mass would not be protected against contributions from the new massive states and fine-tuning would be needed. If the SM is embedded in a supersymmetric theory, the Higgs mass is radiatively stable down to the scale of supersymmetry breaking, which has to be in the vicinity of the electroweak breaking. Nevertheless, the new degrees of freedom would have to be observed in this neighborhood and the results from LHC have been negative for supersymmetry so far. 

Another possibility is that of {\textit{scale invariance}}. In the case that at very high energies the Higgs mass vanishes and the theory, having no dimensionful parameter, is classically scale invariant, no such mass term can arise by radiative corrections at lower energy scales~\cite{Bardeen1995}. Only logarithms multiplying the tree-level terms can arise and the Higgs effective potential will have the form
\be
V_{\text{eff}}(H)\,=\,\lambda\,|H|^4\,+\,C\,|H|^4\,\ln(|H|^2/\mu^2){\label{CW}}
\,.
\ee
However, a dimensionful parameter and, in consequence, physical masses can arise through spontaneous symmetry breaking. The associated emergence of a dimensionful parameter from a scale invariant theory has been termed {\textit{dimensional transmutation}} and its realization through radiative corrections {\textit{Coleman-Weinberg mechanism}}~\cite{Coleman1973}. Minimization of \eqref{CW} gives $\langle H\rangle\,\sim\,\mu\,e^{-\lambda(\mu)/C}$.
Taking the renormalization scale $\mu$ at the UV, we can in principle account for an exponential hierarchy between the electroweak scale $\langle H\rangle$ and the Planck scale $\mu=M_P$~\cite{Foot2010, Salvio2014a, Einhorn2015, Kannike2015}, provided the particle content of the theory is such that the radiative corrections coefficient $C$ is positive. Nevertheless, in the minimal version of the Standard Model the dominant contribution in $C$ comes from the top quark and makes it negative, therefore no hierarchy is generated. This can change if extra bosonic degrees of freedom with sizable couplings are introduced. These could render $C$ positive and sufficiently large in order to generate the desired hierarchy. The extra degrees of freedom could be scalars or gauge bosons, with the latter possibility corresponding to an enlargement of the gauge group.

The observed Higgs mass $M_h=125.09\,\,\GeV$, resulting in $\lambda(M_t)\approx 0.1285$\cite{Sirlin1986,Holthausen2012}, leads to negative values for the Higgs self-coupling $\lambda(\mu)$ above scales of $\mathcal{O}(10^{10}\,\,\GeV)$ which means the vacuum is actually \textit{metastable}~\cite{Degrassi2012, Buttazzo2013, Zoller2014, Branchina2013, Branchina2014, Andreassen2014, Branchina2015, Bezrukov2014, Espinosa2015, Branchina2015a, Bednyakov2015}. This behavior can be avoided in extended versions of the Standard Model, the simplest of which consists in the introduction of a gauge singlet scalar field that couples only to the Higgs field. One way that the extra field affects stability is through the positive contribution to the renormalization group equation of $\lambda$ induced by its portal coupling to the Higgs that tends to counterbalance the negative top quark contribution. Barring the unappealing possibility of introducing an additional \textit{ad hoc} mass scale associated with this singlet, one may incorporate the new field in the classically scale invariant version of the Standard Model~\cite{Meissner2007, Foot2007, Alexander-Nunneley2010, Gabrielli2014, Antipin2014, Sannino2015, Farzinnia2015}. This allows for a second way to affect stability due to the increase of the low-energy value of the Higgs coupling caused by the extra contributions from the vacuum expectation value (vev) of the singlet field.

The new scalar can acquire its vev through dimensional transmutation by a realization of the Coleman-Weinberg mechanism (CW). Loop corrections to its effective potential can generate for it a nonzero vev that is fed to the standard Higgs through the portal coupling. However, as we remarked above discussing \eqref{CW}, the coefficient $C$ of the one-loop effective scalar potential has to be positive. This can be readily achieved if the new field is charged under a new gauge group, since the new gauge boson contributions to $C$ are positive. Thus, introducing an extra dark gauge group~\cite{Hempfling1996, Chang2007a, Iso2009, Holthausen2010, Iso2013, Englert2013, Khoze2013, Khoze2013a, Hambye2013, Carone2013, Heikinheimo2014, Hashimoto2014, Pelaggi2015, Guo2015, Haba2015, Orikasa2015, Plascencia2015} with extra scalar degrees of freedom can in principle lead to a successful realization of CW.

The new sector associated with UV stabilization and low-energy symmetry breaking can also in principle account for dark matter. A characteristic example is provided by dark matter being a vector boson, resulting from the breaking of an extra $SU(2)_X$ gauge symmetry \cite{Hambye2013,Carone2013}, with mass in the $\TeV$ range. Depending on the model, additional dark degrees of freedom may be needed in order to saturate the dark matter relic abundance. These could be additional dark scalars \cite{Endo2015a, Ishiwata2012, Farzinnia2014, Khoze2014}, although fermions are allowed too if their masses are light enough not to invalidate the dark sector CW mechanism\cite{Altmannshofer2015, Benic2014, Benic2015}.

Regarding neutrino masses, we can introduce right-handed neutrinos as singlets~\cite{Dias2006, Foot2007a, Davoudiasl2014, Kannike2014, Lindner2014, Kang2014, Humbert2015, Ahriche2015}, neutral both under the Standard Model as well as the dark gauge group. A mass term for them can arise from a coupling to a total singlet scalar that obtains a vev through its couplings to the dark sector. Thus, neutrinos may acquire masses as a result of a seesaw mechanism driven by the scale of the dark sector. Nevertheless, this low-energy seesaw procedure still requires small Yukawa couplings for the left-handed neutrinos of the order of the corresponding electron Yukawa coupling since the singlet vev, tied to the gauge symmetry breaking, cannot be too large.

In the present paper we have reconsidered an $SU(2)_X$ extension of the classically scale invariant (CSI) version of the Standard model and, in the light of the breaking of the gauge symmetry $SU(2)_L\times U(1)_Y\times SU(2)_X\rightarrow U(1)_{em}$, we address the issues of UV stability, the structure of the multi-Higgs sector, dark matter and neutrino masses. Apart from the right-handed neutrinos and a real scalar singlet, having a coupling to them and to the standard Higgs, the model consists of a dark $SU(2)_X$ sector composed of three dark gauge vectors and a dark isodoublet possessing portal couplings to the standard Higgs and the aforementioned singlet. After symmetry breaking, the extended Higgs sector consists of three states, one to be identified with the standard Higgs and two additional scalars. The resulting dark vectors, being stable and weakly interacting massive particles (WIMPs), can be identified with dark matter. For a suitable range of the free parameters of the model all relevant experimental constraints can be met as well as stability and perturbativity. In addition, the model has a range of definite dark matter predictions which can be either falsified or verified by new limits or observations in near future experiments.

The paper is organized as follows. In the next section, we present the model and analyze the stability of the tree-level potential. We proceed to obtain possible flat directions, setting up the model for the study of symmetry breaking through the Coleman-Weinberg mechanism. We compute the one-loop effective potential and the resulting scalar masses. Subsequently, in Sec. \ref{sec:pheno}, we undertake a phenomenological analysis of the model. We identify one of the predicted scalar states with the observed Higgs boson. Then, we find benchmark sets of values for a minimal subset of the free parameters of the model that correctly reproduce the Higgs boson mass. After that, we scan over the rest of the parameters and obtain masses for the dark gauge bosons, the right-handed neutrinos and one of the scalar bosons, all the while checking that the stability and perturbativity constraints are satisfied. In Sec. \ref{sec:DM}, for the same set of benchmark values, we calculate the dark matter relic density and constrain the masses of the dark gauge bosons from both the observed relic density and the limits set by direct detection experiments. Finally, in Sec. \ref{sec:coclusions} we summarize and conclude.
\section{The model}
\label{model}
In this section we present the model and study its properties. Employing the Coleman-Weinberg mechanism~\cite{Coleman1973} in the Gildener-Weinberg~\cite{Gildener1976} formalism we minimize the tree-level potential and find the flat direction between the vevs of the scalar fields. Then we obtain the tree-level masses of the scalars, one of which (it may be called \textit{darkon}) turns out to be massless due to the flat direction. Including the one-loop potential, we find that radiative corrections become dominant along the flat direction and lift the darkon's mass to values that can be even higher than the masses of the other scalars.
 
\subsection{The tree-level scalar potential}
\label{subsec:potential}

In order to address the open issues discussed in the Introduction we consider the Standard Model in a classically scale invariant (CSI) framework and extend the gauge group with an additional $SU(2)_X$ symmetry~\cite{Hambye2013, Carone2013}. In addition to the new gauge bosons, the dark sector contains a SM-singlet scalar $SU(2)_X$ isodoublet $\Phi$. Aiming at the problem of neutrino mass generation, we also introduce a real scalar $\sigma$, singlet under both the SM and dark gauge groups. Right-handed neutrinos are also included in the standard fashion as total fermionic singlets. The tree-level scalar potential, in terms of the SM Higgs field $H$ and the new scalars $\Phi,\,\sigma$, has the form
\be 
  V_0\, =\,\shh ( H^{\dagger} H )^2 + \sff ( \Phi^{\dagger} \Phi )^2 +  \frac{\ssisi}{4} \sigma^4 - \shf ( H^{\dagger} H ) ( \Phi^{\dagger} \Phi )  - \frac{\sfsi}{2} ( \Phi^{\dagger} \Phi ) \sigma^2 
   + \frac{\shsi}{2} ( H^{\dagger} H ) \sigma^2,\label{tlpot}  
\ee
where we included all possible couplings among scalars and have introduced negative signs for the portal couplings $\lambda_{h\phi}$ and $\lambda_{\phi\sigma}$. Models having all mixing scalar couplings positive, i.e. both the portals to the dark sector $\lambda_{h\phi}$, $\lambda_{\phi\sigma}$ and the observable sector Higgs mixing $\lambda_{h\sigma}$, do not lead to a symmetry breaking flat direction and are not of interest. It is then reasonable to examine models with the portal to the dark sector being negative. Taking the observable mixing $\lambda_{h\sigma}$ also negative is not necessarily interesting because it allows for a flat direction independent of the dark sector. Therefore, we restrict the possible breaking patterns making the above choice of signs which is sufficient for our purposes.

In addition to the scalar potential, the Lagrangian has the following extra Yukawa terms:
\be 
-\mathcal{L}_N = Y^{ij}_{\nu} \bar{L}_i \, i\sigma_2 H^{*} N_j + \text{H.c.} + {Y^{ij}_\sigma} \bar{N}^{c}_i N_j \sigma, 
\label{YukawaTerms}
\ee
where $Y^{ij}_{\nu}$ is the Dirac neutrino Yukawa matrix which couples the left-handed lepton doublet $L_i$ to the SM Higgs doublet $H$ and the right-handed neutrino $N_j$ and $Y^{ij}_\sigma$ is the right-handed Majorana neutrino Yukawa matrix which will be assumed diagonal.

Considering the unitary gauge and the symmetry breaking pattern $SU(2)_L\times U(1)_Y\times SU(2)_X\rightarrow\,SU(2)_L\times U(1)_Y\rightarrow\,U(1)_{em}$, we may replace the scalar doublets by
\be 
H = \frac{1}{\sqrt{2}}\left(\begin{array}{c}
0\\
h
\end{array}\right), \quad  \Phi = \frac{1}{\sqrt{2}}\left(\begin{array}{c}
0\\
\phi
\end{array}\right)\,.
\label{UNIFIELDS}
\ee
Then, the tree-level potential takes the form
\be 
V_{0} (h,\phi,\sigma) = \frac{\shh}{4} h^4 + \frac{\sff}{4} \phi^4 + \frac{\ssisi}{4} \sigma^4 - \frac{\shf}{4} h^2 \phi^2  - \frac{\sfsi}{4} \phi^2 \sigma^2  + \frac{\shsi}{4} h^2 \sigma^2.
\label{treelevelpotential}
\ee
The scalar potential is bounded from below if the matrix
\be 
\mathcal{A}\,=\, \frac{1}{8}\,\left(\begin{array}{ccc}2\lambda_h&-\lambda_{h\phi}&\lambda_{h\sigma}\\
 \,&\,&\,\\
 -\lambda_{h\phi}&2\lambda_{\phi}&-\lambda_{\phi\sigma}\\
 \,&\,&\,\\
 \lambda_{h\sigma}&-\lambda_{\phi\sigma}&2\lambda_{\sigma}
 \end{array}\right)
\ee
is {\textit{copositive}}, i.e. such that $\eta_a \mathcal{A}_{ab}\eta_b$ is positive for non-negative vectors in the basis $(h^2,\,\phi^2,\,\sigma^2)$. It can be shown\cite{Hadeler1983,Ping1993,Chang1994,Kannike2012,Chakrabortty2014} that this is equivalent to the conditions
\be
\lambda_h\,\geq\,0,\,\lambda_{\phi}\,\geq\,0,\,\lambda_{\sigma}\,\geq\, 0 \label{stabilityconditions1}
\ee
\be\frac{\lambda_{h\phi}}{2\sqrt{\lambda_h\lambda_{\phi}}}\,\leq\,1,\,\,\,\frac{{-\lambda_{h\sigma}}}{ 2\sqrt{\lambda_h\lambda_{\sigma}}}\,\leq\,1,\,\,\frac{\lambda_{\phi\sigma}}{ 2\sqrt{\lambda_{\phi}\lambda_{\sigma}}}\,\leq\,1 \label{stabilityconditions2}
\ee
\be
\left[ 2 \left(1-\frac{\lambda_{h\phi}}{2\sqrt{\lambda_h\lambda_{\phi}}}\right)\left(1+\frac{\lambda_{h\sigma}}{2\sqrt{\lambda_h\lambda_{\sigma}}}\right)\left(1-\frac{\lambda_{\phi\sigma}}{2\sqrt{\lambda_{\phi}\lambda_{\sigma}}}\right) \right]^{1/2}\,\geq\,-1\,+\,\frac{\lambda_{\phi\sigma}}{2\sqrt{\lambda_{\phi}\lambda_{\sigma}}}\,+\frac{\lambda_{h\phi}}{2\sqrt{\lambda_h\lambda_{\phi}}}\,-\frac{\lambda_{h\sigma}}{2\sqrt{\lambda_h\lambda_{\sigma}}} \label{stabilityconditions3}
\ee
Note that the last condition is equivalent to \textit{either} of the following statements: 
\ba
&\frac{\lambda_{h\phi}}{2\sqrt{\lambda_h\lambda_{\phi}}}\,-\frac{{\lambda_{h\sigma}}}{2\sqrt{\lambda_h\lambda_{\sigma}}}\,+\,\frac{\lambda_{\phi\sigma}}{2\sqrt{\lambda_{\phi}\lambda_{\sigma}}}\,\leq\,1 , \label{stabilityconditions4}\\
&\det \mathcal{A}\,=\, \lambda_h\lambda_{\phi}\lambda_{\sigma}\,-\frac{1}{4}\left(\lambda_{h\phi}^2\lambda_{\sigma}+\lambda_{h\sigma}^2\lambda_{\phi}+ \lambda_{\phi\sigma}^2\lambda_h\right)\,+ \frac{1}{4}\lambda_{h\phi}\lambda_{h\sigma}{\lambda_{\phi\sigma}}\,\geq\,0. \label{stabilityconditions5}
\ea
Therefore, vacuum stability requires the validity of the above conditions to hold at all energies up to $M_{P}$. In order to study the flat directions of the tree-level potential we may parametrize the scalar fields as
\be 
h = \varphi N_1, 
\quad 
\phi = \varphi N_2, 
\quad 
\sigma = \varphi N_3,
\ee
with $N_i$ a unit vector in the three-dimensional field space. Then, the tree-level potential attains the form
\be 
V_0=\frac{\varphi^4}{4}\left[ \shh N^4_1 + \sff N^4_2 + \ssisi N^4_3 - \shf N^2_1 N^2_2 + \shsi N^2_1 N^2_3 - \sfsi N^2_2 N^2_3 \right].
\ee
The condition for an extremum along a particular direction $N_i=n_i$ is \cite{Gildener1976}
\be 
\left.\frac{\partial V_0}{\partial N_i}\right|_{\bf{n}}\,=\,V_0(\mathbf{n})\,=\,0\,.{\label{MINI}}
\ee
Then, the equations giving the symmetry breaking direction are
\ba 
2\shh n_1^2 &=& \shf n_2^2 - \shsi n_3^2 \label{FlatDirectionA} \\
2\sff n_2^2 &=& \shf n_1^2 + \sfsi n_3^2 \label{FlatDirectionB} \\
2\ssisi n_3^2 &=& \sfsi n_2^2 - \shsi n_1^2 \label{FlatDirectionC} 
\ea
\be
\shh n_1^4 + \sff n_2^4 + \ssisi n_3^4 - \shf n_1^2 n_2^2 - \sfsi n_2^2 n_3^2 + \shsi n_1^2 n_3^2 = 0. \label{FlatDirectionD}
\ee
The solution of these equations in terms of the scalar couplings is
\be 
\begin{array}{l}
n_1^2\,=\,\frac{4 \lambda_{\sigma} \lambda_{\phi} - \lambda^2_{\phi\sigma}}{2 \lambda_{\sigma} \left( 2 \lambda_{\phi} + \lambda_{h \phi} \right) + \lambda_{\phi \sigma} \left( \lambda_{h \phi} - \lambda_{\phi \sigma} \right) - \lambda_{h\sigma} \left( 2 \lambda_\phi + \lambda_{\phi\sigma} \right) }\\
\,\\
n_2^2\,=\,\frac{2 \lambda_\sigma \lambda_{h\phi} - \lambda_{h\sigma} \lambda_{\phi\sigma}}{2 \lambda_{\sigma} \left( 2 \lambda_{\phi} + \lambda_{h \phi} \right) + \lambda_{\phi \sigma} \left( \lambda_{h \phi} - \lambda_{\phi \sigma} \right) - \lambda_{h\sigma} \left( 2 \lambda_\phi + \lambda_{\phi\sigma} \right)}\\
\,\\
n_3^2\,=\,\frac{\lambda_{h\phi} \lambda_{\phi\sigma} - 2 \lambda_\phi \lambda_{h\sigma}}{2 \lambda_{\sigma} \left( 2 \lambda_{\phi} + \lambda_{h \phi} \right) + \lambda_{\phi \sigma} \left( \lambda_{h \phi} - \lambda_{\phi \sigma} \right) - \lambda_{h\sigma} \left( 2 \lambda_\phi + \lambda_{\phi\sigma} \right)}
\end{array}
\ee
Note that $n_1^2 + n_2^2 + n_3^2 = 1$.
\subsection{The scalar masses}
\label{subsec:scalar masses}
Assuming spontaneous breaking of the gauge and the scale symmetry via the Coleman-Weinberg mechanism, we can write the shifted scalar fields as
\be
h\,=\,(\varphi+v)\,n_1, \quad \phi\,=\,(\varphi+v)\,n_2, \quad\sigma\,=\,(\varphi+v)\,n_3 .
\ee
The individual vevs are
\be 
\langle h\rangle\,\equiv\,v_h=v\,n_1,\,\,\langle\phi\rangle\,\equiv\,v_{\phi}=vn_2,\,\,\langle\sigma\rangle\,\equiv\,v_{\sigma}=vn_3\,.
\ee
From the shifted tree-level potential we can read off the scalar mass matrix
\be 
\mathcal{M}_0^2 = \upsilon^2 
\left( \begin{array}{ccc}
2\shh n_1^2    & -n_1 n_2 \shf  & +n_1 n_3 \shsi  \\
-n_1 n_2 \shf  & 2\sff n_2^2    & -n_2 n_3 \sfsi  \\
+n_1 n_3 \shsi & -n_2 n_3 \sfsi &  2\ssisi n_3^2
\end{array}
\right)\label{massmatrix}
\ee
in the $(h, \phi, \sigma)$ basis. We can now set up the diagonalization of the mass matrix \eqref{massmatrix} by introducing a general rotation in terms of three parametric angles,
\be 
{\cal{R}}\,{\cal{M}}_0^2\,{\cal{R}}^{-1}\,=\,{\cal{M}}_{d}^2,
\ee
with the rotation matrix ${\cal{R}}^{-1}$ given by
\be 
{\cal{R}}^{-1}\,=\,\left( \begin{array}{ccc}
\cos\alpha\cos\beta  & \sin\alpha & \cos\alpha\sin\beta  \\
-\cos\beta\cos\gamma\sin\alpha + \sin\beta\sin\gamma      & \cos\alpha\cos\gamma   & -\cos\gamma\sin\alpha\sin\beta - \cos\beta\sin\gamma \\
-\cos\gamma\sin\beta - \cos\beta\sin\alpha\sin\gamma  & \cos\alpha\sin\gamma & \cos\beta\cos\gamma - \sin\alpha\sin\beta\sin\gamma
\end{array}
\right)\label{rotationmatrix}
\ee
and
\be
\left(\begin{array}{c}
h \\
\phi \\
\sigma 
\end{array}
\right)
=  
\left(\begin{array}{ccc}
.&.&.\\
.&{\cal{R}}^{-1}&.\\
.&.&.
\end{array}\right)
\left(
\begin{array}{c}
h_1 \\
h_2 \\
h_3
\end{array}
\right).
\ee
Next, we choose two of the above three angles in the rotation matrix to parametrize the total vev $v$ direction according to
\be 
\begin{array}{l} 
v_h = v \sin\alpha = v n_1  \\
v_\phi = v  \cos\alpha\cos\gamma = v n_2 \\
v_\sigma = v \cos\alpha\sin\gamma = v n_3.
\end{array}
\label{parametrize}
\ee
Then, ${\cal{M}}_d^2$ is diagonal, provided that the following relations are satisfied:
\be
\begin{array}{l}
\tan^2\alpha = \frac{v_h^2}{v_\phi^2 + v_\sigma^2} = \frac{4 \sff \ssisi - \sfsi^2}{2 \left( \ssisi \shf - \sff \shsi \right) + \sfsi \left( \shf - \shsi \right)}\\
\,\\
\tan^2\gamma = \frac{v_\sigma^2}{v_\phi^2} = \frac{2\shh \sfsi - \shf \shsi}{4 \shh \ssisi - \shsi^2} \\
\,\\
\tan2\beta = \frac{v_h v_\phi v_\sigma v \left( \shsi + \shf \right)}{\left( \sff + \ssisi + \sfsi \right) v_\phi^2 v_\sigma^2 - \shh v_h^2 v^2}.
\end{array} \label{angles}
\ee
The resulting mass eigenvalues are
\be 
\begin{split}
M^2_{h_1}/2  =&   \shh v_h^2 \cos^2\alpha \cos^2\beta + \sff v_\phi^2 \left( \cos\beta \cos\gamma \sin\alpha - \sin\beta \sin\gamma \right)^2 \\ 
+& \ssisi v_\sigma^2 \left( \cos\gamma \sin\beta + \cos\beta \sin\alpha \sin\gamma \right)^2 \\
+& \shf v_h v_\phi \cos\alpha \cos\beta \left( \cos\beta \cos\gamma \sin\alpha - \sin\beta \sin\gamma \right) \\
-& \sfsi v_\phi v_\sigma \left( \cos\beta \cos\gamma \sin\alpha - \sin\beta \sin\gamma \right) \left( \cos\gamma \sin\beta + \cos\beta \sin\alpha \sin\gamma \right) \\
-& \shsi v_h v_\sigma \cos\alpha \cos\beta \left( \cos\gamma \sin\beta + \cos\beta \sin\alpha \sin\gamma \right)
\end{split} \label{massh1}
\ee
\be 
M^2_{h_2} = 0 \label{massh2}
\ee
\be 
\begin{split}
M^2_{h_3}/2 \, =& \,  \shh v_h^2 \cos^2\alpha \sin^2\beta + \sff v_\phi^2 \left( \sin\beta \cos\gamma \sin\alpha + \cos\beta \sin\gamma \right)^2 \\ 
+& \ssisi v_\sigma^2 \left( \cos\gamma \cos\beta - \sin\beta \sin\alpha \sin\gamma \right)^2 \\
+& \shf v_h v_\phi \cos\alpha \sin\beta \left( \sin\beta \cos\gamma \sin\alpha + \cos\beta \sin\gamma \right) \\
+& \sfsi v_\phi v_\sigma \left( \sin\beta \cos\gamma \sin\alpha + \cos\beta \sin\gamma \right) \left( \cos\gamma \cos\beta - \sin\beta \sin\alpha \sin\gamma \right) \\
+& \shsi v_h v_\sigma \cos\alpha \sin\beta \left( \cos\gamma \cos\beta - \sin\beta \sin\alpha \sin\gamma \right)
\end{split} \label{massh3}
\ee
As expected, one of these masses turns out to be zero at tree level. Regarding the rest, $M_{h_1}$ and $M_{h_3}$ are ultimately functions of the overall vev $v$ and the scalar couplings. Since their exact, analytic expressions are not necessary, we leave them as they stand.
\subsection{Neutrinos}
\label{subsec:neutrinos}
One of the defining properties of the present model is that it incorporates the appropriate structure for massive Majorana neutrinos. Right-handed neutrinos in three families are introduced as singlets of both the Standard Model and the $SU(2)_X$ dark sector. They obtain their mass as a result of broken scale invariance through their coupling to the singlet $\sigma$ that mediates between the two sectors and obtains a nonzero vev. The Yukawa terms \eqref{YukawaTerms} that give rise to neutrino masses are
\be 
\frac{Y^{ij}_{\nu}}{\sqrt{2}} v_h \nu_i \, i\sigma_2  N_j + \text{H.c.} + {Y^{ij}_\sigma}v_{\sigma} \bar{N}^{\;c}_i N_j \,.
\ee
The neutrino mass matrix, being of the seesaw type, can lead to the desired scale of $\mathcal{O}(0.1\,\,\eV)$ for the left-handed neutrino masses. Thus, in a $(\nu_s,\,N_i^c)$ basis, we have
\be
 \left(\begin{array}{cc}
0\,&\,\frac{Y^{ij}_{\nu}}{\sqrt{2}} v_h \\
\,&\,\\
\frac{Y^{ij}_{\nu}}{\sqrt{2}} v_h \,&\,{Y^{ij}_\sigma}v_{\sigma}
\end{array}\right).
\ee
Assuming $Y^{ij}_{\nu}\,v_h$ to be no more than the lightest charged lepton mass, namely $\mathcal{O}(10^{-4} \,\, \GeV)$, and taking characteristic values $v_{\sigma}\sim \mathcal{O}(1 \,\, \TeV)$ and $Y_{\sigma}\sim \mathcal{O}(0.1)$, we have $Y_{\nu}\,v_h\, \ll \,Y_{\sigma}\,v_{\sigma}$ and we arrive at approximate eigenvalues
\be 
M_N\,\approx\,{Y^{ij}_\sigma}v_{\sigma},\quad m_{\nu}\,\approx\,\frac{v_h^2}{4v_{\sigma}}\,Y_{\nu}^{(ik)}\,\left(Y_{\sigma}^{-1}\right)^{(k\ell )}\,Y_{\nu}^{(\ell j)},
\ee
with $M_N\,\sim\,\mathcal{O}(100 \,\, \GeV)$ and $m_{\nu}\,\sim\, \mathcal{O}(0.1 \,\, \eV)$. As we will see next, the right-handed neutrino mass scale is related to the masses of the rest of the particles and cannot take arbitrary values.
\subsection{The one-loop potential}
\label{subsec:1-loop}
Now, let us consider the full one-loop potential. Following the Gildener-Weinberg approach~\cite{Gildener1976} to symmetry breaking we have
\be 
\left. \frac{\partial}{\partial\Phi_i}(V_0+V_1)\right|_{\mathbf{\Phi}=v(\mathbf{n}+\delta\mathbf{n})}\,=\,0
\ee
or, expanding and using \eqref{MINI}, 
\be 
0\,=\,\left.\frac{\partial V_0}{\partial N_i}\right|_{\mathbf{N}=\mathbf{n}}\,+\,\delta n_j\left.\frac{\partial^2V_0}{\partial N_i\partial N_j}\right|_{\mathbf{N}=\mathbf{n}}\,+\,\left.\frac{\partial V_1}{\partial N_i}\right|_{\mathbf{N}=\mathbf{n}}\,=\,\delta n_j\left.\frac{\partial^2V_0}{\partial N_i\partial N_j}\right|_{\mathbf{N}=\mathbf{n}}\,+\,\left.\frac{\partial V_1}{\partial N_i}\right|_{\mathbf{N}=\mathbf{n}}. 
\ee 
Contracting with $n_i$, we obtain\footnote{The tree-level scale invariance enforces
$$\Phi_i\frac{\partial V_0}{\partial\Phi_i}\,=\,4V_0\,\Longrightarrow \left.\Phi_i\frac{\partial^2V_0}{\partial\Phi_i\partial\Phi_j}\right|_{\mathbf{\Phi}=v\mathbf{n}}\,=\,\left.\frac{\partial V_0}{\partial\Phi_j}\right|_{\mathbf{\Phi}=v\mathbf{n}}=0.$$}
\be 
n_i\delta n_j\left.\frac{\partial^2V_0}{\partial N_i\partial N_j}\right|_{\mathbf{N}=\mathbf{n}}\,+\,n_i\left.\frac{\partial V_1}{\partial N_i}\right|_{\mathbf{N}=\mathbf{n}}\,=\,n_i\left.\frac{\partial V_1}{\partial N_i}\right|_{\mathbf{N}=\mathbf{n}}\,=\,0\,. 
\ee
The last statement is equivalent to
\be 
\left.\frac{\partial V_1(\mathbf{n}\varphi)}{\partial\varphi}\right|_{\varphi=v}\,=\,0. 
\label{MINIFULL}
\ee
In this approach the couplings of the tree-level potential are assumed to depend on the renormalization scale $\mu$ and the tree-level minimization condition to be realized at a particular value $\mu=\Lambda$. Thus, along the minimum flat direction at the scale $\Lambda$ the one-loop effective potential has the form
\be 
V_1(\mathbf{n}\varphi) = A \varphi^4 + B \varphi^4 \log\frac{\varphi^2}{\Lambda^2}.
\ee
The coefficients $A$ and $B$ are dimensionless parameters and are given in the $\overline{MS}$ scheme by
\be 
\begin{split}
A &= \frac{1}{64\pi^2 \upsilon^4} \left[ \sum_{i=1,3} M^4_{h_i} \left( -\frac{3}{2} + \log \frac{M^2_{i}}{\upsilon^2} \right) +6 M^4_W \left( -\frac{5}{6} + \log \frac{M^2_W}{\upsilon^2}  \right) + 3 M^4_Z \left( -\frac{5}{6} + \log \frac{M^2_Z}{\upsilon^2} \right) \right.
  \\ & \left. + 9 M^4_X \left( -\frac{5}{6} + \log \frac{M^2_X}{\upsilon^2}  \right) - 12 M^4_t \left( -1 + \log \frac{M^2_t}{\upsilon^2}  \right) - 2 \sum^3_{i=1} M^4_{N_i} \left( -1 + \log \frac{M^2_{N_i}}{\upsilon^2}  \right) \right],
\end{split}
\ee
\be 
B = \frac{1}{64\pi^2 \upsilon^4} \left(\sum_{i=1,3} M^4_{h_i}  +6 M^4_W  + 3 M^4_Z 
  + 9 M^4_X  - 12 M^4_t  - 2 \sum^3_{i=1} M^4_{N_i}  \right).
\ee
The condition \eqref{MINIFULL} gives
\be 
\log\left( \frac{\upsilon}{\Lambda} \right) = -\frac{1}{4} - \frac{A}{2B} 
\label{LAMBDA}.
\ee
Thus the one-loop effective potential becomes
\be 
V_1(\mathbf{n}\varphi) = B \varphi^4 \left[ \log\frac{\varphi^2}{\upsilon^2} - \frac{1}{2} \right].
\label{onelooppotential}
\ee
The pseudo-Goldstone boson (darkon) mass is now shifted from zero to
\be 
M^2_{h_2} = \left. \frac{\D^2 V_1(\mathbf{n}\varphi)}{\D \varphi^2}\right\vert_{\varphi=\upsilon} = \frac{1}{8\pi^2 \upsilon^2} \left(M^4_{h_1}+M^4_{h_3}  +6 M^4_W  + 3 M^4_Z 
  + 9 M^4_X  - 12 M^4_t  - 6M^4_{N}  \right), 
\label{PLBMASS}
\ee
where we have assumed for simplicity that all right-handed neutrinos are degenerate in mass. Note that the right-handed neutrino contribution to \eqref{PLBMASS}, as fermionic, enters with a minus sign. As a result, $M_N$ cannot be too large. In what follows we shall identify the state $h_1$ with the observed Higgs of $125.09\,\,\GeV$ and choose a higher value for the $h_3$ mass. The $h_2$ state, although massless at tree level, can have any mass with respect to the other scalars due to the sizable one-loop correction. Note that radiative corrections to the tree-level masses of $h_1$ and $h_3$ are small enough to ignore to a first approximation. 

\section{Phenomenological analysis}
\label{sec:pheno}
In this section we present an analysis of the phenomenological viability of the model, taking into account theoretical and experimental constraints. Our procedure in broad terms will be as follows: First, we choose values from a subset of the free parameters of the model (i.e. $v_{\phi},\,v_{\sigma}$ and some of the scalar couplings), appropriate to fix the mass $M_{h_1}$ to the experimental value of $125.09\,\,\GeV$. Then the $M_{h_3}$ mass is automatically obtained. In order to calculate the darkon mass $M_{h_2}$ we scan over the two remaining unknown masses $M_X$ and $M_N$ in \eqref{PLBMASS}, while checking that the stability and perturbativity conditions are satisfied. Finally, we calculate the total decay rates of all the scalar bosons and compare the one corresponding to the Higgs boson with the bounds set by LHC.
\subsection{Theoretical constraints}
\label{subsec:theoryconst}

The tree-level potential \eqref{treelevelpotential} and the one-loop effective potential \eqref{onelooppotential} have to be bounded from below for the vacuum to be stable. For this to be valid, the stability conditions \eqref{stabilityconditions1}-\eqref{stabilityconditions5} need to hold for all energies up to the Planck scale $(M_{\text{P}} = 1.22 \times 10^{19} \,\, \GeV)$ as well as the positivity condition $B>0$ has to be satisfied. The latter translates to
\be 
M^4_{h_3} + 9 M^4_X - 6 M^4_{N} >  12 M^4_t - 6 M^4_W - 3 M^4_Z - M^4_{h_1}
\label{BPOSITIVE}
\ee
or
\be 
M^4_{h_3} + 9 M^4_X - 6 M^4_{N} > \left( 317.26 \,\, \GeV \right)^4 ,
\label{massesinequality}
\ee
where we used the values $M_t = 173.34\,\,\GeV$~\cite{2014}, $M_W = 80.384\,\,\GeV$ and $M_Z = 91.1876\,\,\GeV$. The above inequality implies that the masses of the extra gauge bosons $M_X$ have to be in general larger than the masses of the right-handed neutrinos $M_N$, unless the scalar boson mass $M_{h_3}$ is considerably larger than the right-hand side of \eqref{massesinequality}.

Another constraint arises from the requirement that the model must remain perturbative all the way up to $M_{\text{P}}$. This can be achieved by demanding that all couplings are bounded,
\be 
\text{all couplings} < 2\,\pi.
\label{perturbativity}
\ee
To determine how the couplings of the model vary with energy, we need to solve the renormalization group equations (RGEs). We present the two-loop gauge and one-loop Yukawa and scalar RGEs below (however in our numerical analysis we use the full two-loop RGEs for all the couplings, computed using Refs.~\cite{Machacek1983a, Machacek1984a, Machacek1985a, Luo2003a, Luo2003b}):
{\allowdisplaybreaks \begin{align} 
\beta_{g_1} & =  
\frac{41}{10} g_{1}^{3} + \frac{1}{(4\pi)^2} \frac{1}{50} g_{1}^{3} \Big( 199 g_{1}^{2} + 135 g_{2}^{2} + 440 g_{3}^{2} -85 y_t^2 \Big) \\ 
\beta_{g_2} & =  -\frac{19}{6} g_{2}^{3} + \frac{1}{(4\pi)^2} \frac{1}{30} g_{2}^{3} \Big( 27 g_{1}^{2} + 175 g_{2}^{2} + 360 g_{3}^{2} -45 y_t^2 \Big)  \\  
\beta_{g_3} & =  -7 g_{3}^{3}  + \frac{1}{(4\pi)^2}\frac{1}{10} g_{3}^{3} \Big( 11 g_{1}^{2} + 45 g_{2}^{2} - 260 g_{3}^{2} - 20 y_t^2  \Big)\\  
\beta_{g_X} & =  -\frac{43}{6} g_{X}^{3}  -\frac{1}{(4\pi)^2}\frac{259}{6} g_{X}^{5} \\
\beta_{y_t} & = y_t \left( \frac{9}{2} y_t^2 -\frac{17}{20} g_{1}^{2} -\frac{9}{4} g_{2}^{2} -  8 g_{3}^{2} \right) \\ 
\beta_{Y_\sigma} & =  
4 \, Y_\sigma \, \mbox{Tr}\Big({Y_\sigma  Y_\sigma^*}\Big)  + 12 \, {Y_\sigma  Y_\sigma^*  Y_\sigma} \\
\beta_{\lambda_h} & =  -6 y_t^4 +24 \lambda_{h}^{2}  +\lambda_h \left( 12 y_t^2 -\frac{9}{5} g_{1}^{2}  -9 g_{2}^{2} \right) +\frac{27}{200} g_{1}^{4} +\frac{9}{20} g_{1}^{2} g_{2}^{2} +\frac{9}{8} g_{2}^{4}  +2 \lambda_{h\phi}^{2} +\frac{1}{2} \lambda_{h\sigma}^{2}  \\ 
\beta_{\lambda_\phi} & =  \frac{9}{8} g_{X}^{4} -9 g_{X}^{2} \lambda_\phi +
 24 \lambda_{\phi}^{2}  + 2 \lambda_{h\phi}^{2}   + \frac{1}{2} \lambda_{\phi\sigma}^{2}   \label{lambdaphiRGE}  \\ 
\beta_{\lambda_\sigma} & =  
-64 \mbox{Tr}\Big({Y_\sigma  Y_\sigma^*  Y_\sigma  Y_\sigma^*}\Big) + 16 \lambda_\sigma \mbox{Tr}\Big({Y_\sigma  Y_\sigma^*}\Big)  + 18 \lambda_{\sigma}^{2}  + 2 \lambda_{h\sigma}^{2}  + 2 \lambda_{\phi\sigma}^{2} \label{lambdasigmaRGE}  \\ 
\beta_{\lambda_{h\phi}} & =  \lambda_{h\phi} \left(6 y_t^2 + 12 \lambda_h + 12 \lambda_\phi -4 \lambda_{h\phi}    -\frac{9}{10} g_{1}^{2} -\frac{9}{2} g_{2}^{2} -\frac{9}{2} g_{X}^{2}   \right)
  + \lambda_{h\sigma} \lambda_{\phi\sigma}  \\   
\beta_{\lambda_{\phi\sigma}} & =  \lambda_{\phi\sigma} \left( 8 \mbox{Tr}\Big({Y_\sigma  Y_\sigma^*}\Big) + 12 \lambda_\phi + 6 \lambda_\sigma  -4 \lambda_{\phi\sigma}   -\frac{9}{2} g_{X}^{2} \right)
 +4 \lambda_{h\sigma} \lambda_{h\phi} \\  
\beta_{\lambda_{h\sigma}} & =  \lambda_{h\sigma} \left( 6 y_t^2 +8 \mbox{Tr}\Big({Y_\sigma  Y_\sigma^*}\Big) +12 \lambda_h +6 \lambda_\sigma  + 4 \lambda_{h\sigma} -\frac{9}{10} g_{1}^{2}  -\frac{9}{2} g_{2}^{2}  \right)
    +4 \lambda_{h\phi} \lambda_{\phi\sigma},
\end{align} }
where we defined $\beta_\kappa \equiv \left( 4 \pi \right)^2 \frac{d \kappa}{d \ln \mu}$. 

In order to solve the RGEs we have to specify the boundary conditions for the couplings. For the SM gauge couplings and the top quark Yukawa coupling we use the NNLO values at $M_t$~\cite{Oda2015,Buttazzo2013}:

\begin{align}
g_1(\mu=M_t) &= \sqrt{\frac{5}{3}} \left( 0.35830 + 0.00011 \left( \frac{M_t}{\GeV} - 173.34 \right) 
			- 0.00020\left(  \frac{M_W - 80.384{\GeV}}{0.014{\GeV}} \right) \right) \\
g_2(\mu=M_t) &= 0.64779 + 0.00004 \left( \frac{M_t}{\GeV} - 173.34 \right) 
			+ 0.00011 \left( \frac{M_W - 80.384{\GeV}}{0.014{\GeV}} \right)  \\			
g_3(\mu=M_t) &= 1.1666 + 0.00314 \left( \frac{\alpha_s(M_Z) - 0.1184}{0.0007} \right) 
			-  0.00046 \left( \frac{M_t}{\GeV} - 173.34 \right)  \\
y_t(\mu=M_t) &= 0.93690 + 0.00556 \left( \frac{M_t}{\GeV} - 173.34 \right) 
			- 0.00042 \left( \frac{\alpha_s(M_Z) - 0.1184}{0.0007} \right). 
\end{align}
In our numerical analysis we use as inputs the central values $\alpha_s(M_Z) = 0.1184$, $M_W = 80.384\,\, \GeV$ and $M_t = 173.34\,\, \GeV$. For the right-handed neutrino Yukawa coupling $Y_\sigma$ and dark gauge coupling $g_X$ we define $Y_\sigma(M_N) = M_N/v_\sigma$ and $g_X(M_X) = 2\,M_X/v_\phi$ respectively. Finally, to define the scalar couplings $\lambda_i$, we consider their values at the renormalization scale $\Lambda$ determined by \eqref{LAMBDA}, where the one-loop effective potential is minimized~\cite{Farzinnia2013}. 

A few comments are in order regarding the behavior of the running couplings. First of all, the $SU(2)_X$ gauge coupling $g_X$ decreases at higher energies, being asymptotically free, in a similar fashion to the $SU(2)_L$ gauge coupling $g_2$. On the other hand, the RGE of the right-handed neutrino Yukawa coupling $Y_\sigma$ has a positive sign and forces $Y_\sigma$ to increase with energy until it potentially reaches a Landau pole. As it turns out, this can be avoided if $Y_\sigma(M_N) \lesssim 0.35$. The Higgs self-coupling $\lambda_h$ generally behaves like the corresponding one in the SM. There, $\lambda_h$ drops fast at increasing energy due to the large negative contribution from the top Yukawa coupling $y_t$, crosses zero at some point and then becomes nearly constant up to the Planck scale. Nevertheless, in our case, we have the freedom to choose a starting value for $\lambda_h$ such that it remains positive inside the whole energy range under consideration. The self-coupling of the singlet scalar $\lambda_\sigma$ depends highly on $Y_\sigma$ and it too can reach a Landau pole unless $Y_\sigma(M_N) \lesssim 0.31$. Therefore $Y_\sigma$ is further constrained. Now, the dark scalar self-coupling $\lambda_\phi$ generally increases with energy, driven mainly by the first term in \eqref{lambdaphiRGE}. A Landau pole is avoided if we have $g_X(M_X) \lesssim 2.51$. Finally, the scalar portal couplings $(\shf,\sfsi,\shsi)$ are mainly multiplicatively renormalized and, as it turns out, they do not run much if we choose initial values that are small enough.

The model contains many free parameters, therefore we need to restrict or fix most of them. The initial eight dimensionless free parameters\footnote{We have assumed that the neutrino Dirac Yukawa coupling takes up values in the neighborhood of the corresponding electron Yukawa coupling.} $\lambda_h,\,\lambda_{\phi},\,\lambda_{\sigma},\,\lambda_{h\phi},\,\lambda_{h\sigma},\,\lambda_{\phi\sigma},\,g_X,\,Y_{\sigma}$ are reduced to six after imposing the experimental values on $v_h$ and $M_{h_1}$. Whether we use the dimensionless scalar couplings or, alternatively, the vevs through the minimization conditions \eqref{FlatDirectionA}-\eqref{FlatDirectionC} as input parameters, is a matter of choice. A six-dimensional parameter space is not easily managed in its full generality. So, we propose to proceed in the following way: We leave $g_X$ and $Y_{\sigma}$ free and list characteristic (benchmark) values for the scalar couplings (at $\Lambda$) and the vevs $v_{\sigma}$ and $v_{\phi}$ that reproduce the measured Higgs mass $M_{h_1} = 125.09\,\, \GeV$. These are shown in Table \ref{table:parameters} where we also show the value for the mass of $h_3$ that we obtain.

\vspace{0.6cm}
\begin{table}[H]
\centering
\begin{tabular}{ | c | c | c | c | c | c | c | c | c | c | c |}
\hline 
Set & $v_h[\GeV]$ & $v_\phi[\GeV]$ & $v_\sigma[\GeV]$ & $\shh (\Lambda)$  & $\sff (\Lambda)$ & $\ssisi (\Lambda)$  & $\shf (\Lambda)$ & $\sfsi (\Lambda)$ & $\shsi (\Lambda)$ & $M_{h_3}[\GeV]$ \\ \hline
A & $246$ & $2112$ & $770$ & $0.1276$ & $0.004$  & $0.2257$ & $0.0036$ & $0.06$ & $0.001$ & $550.62$ \\ 
\hline
B & $246$ & $3245$ & $1470$ & $0.1285$ & $0.0005$  & $0.0122$ & $0.0015$ & $0.005$ & $0.0001$ & $251.93$\\
\hline 
C & $246$ & $4513$ & $2181$ & $0.1287$ & $0.0035$  & $0.0642$ & $0.001$ & $0.03$ & $0.001$ & $868.15$\\
\hline  
\end{tabular}
\caption{Benchmark sets of values for the model parameters able to reproduce the observed Higgs boson mass $M_{h_1}=125.09\,\, \GeV$.}
\label{table:parameters}
\end{table}

\begin{figure}[H]
\centering
{\includegraphics[width=11.5cm]{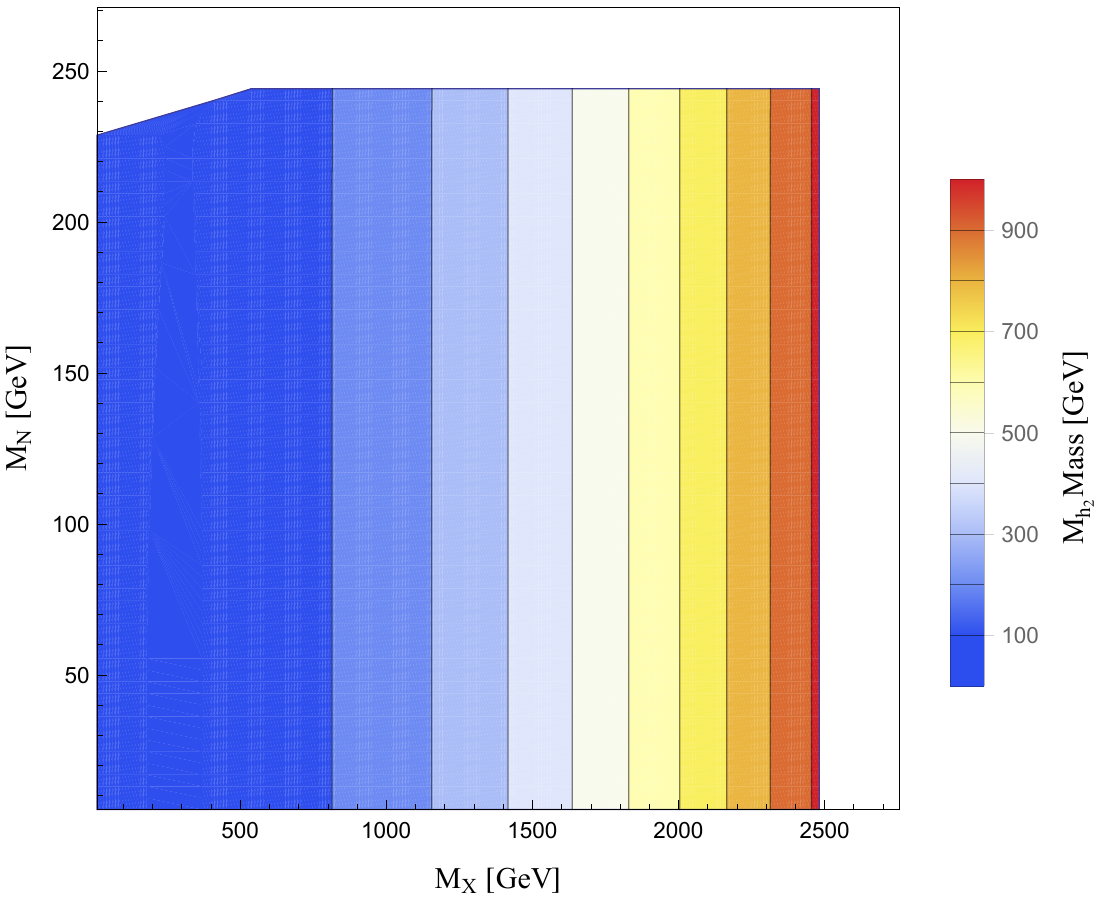}}
\caption{(color online). Parameter space scan in the plane $(g_X,Y_\sigma)$, taking into account constraints from stability and perturbativity. The color coding signifies the mass of the darkon $M_{h_2}$.}
\label{fig:Mh2ContourPlot}
\end{figure}

Not all these sets are compatible with the stability of the potential. For example, with the values in set B the stability condition \eqref{stabilityconditions5} is violated. Choosing the first set of values ($A$), we scan over the remaining two free parameters $g_X$ and $Y_\sigma$ and obtain the darkon mass $M_{h_2}$ contours shown in Fig. \ref{fig:Mh2ContourPlot}, while checking that the stability conditions \eqref{stabilityconditions1}-\eqref{stabilityconditions5} are satisfied, that the one-loop potential is bounded from below \eqref{massesinequality} and that all the couplings remain perturbative up to the Planck scale \eqref{perturbativity}.

\vspace{0.6cm}

\begin{figure}[H]
\centering
{\includegraphics[width=11.5cm]{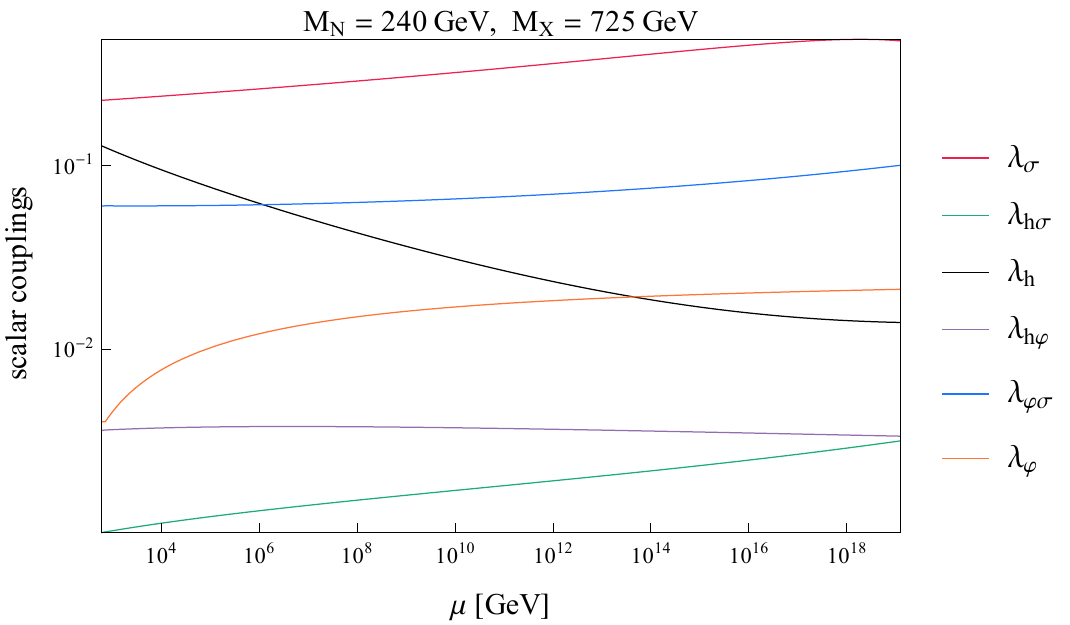}}
\caption{(color online). The RG evolution of the scalar couplings at two-loop order for $M_N = 240\,\, \GeV$ and $M_X = 725\,\, \GeV$.}
\label{fig:RGEsPlot}
\end{figure}

We also present the running of the scalar couplings in Fig. \ref{fig:RGEsPlot}, again for the values of set A in Table \ref{table:parameters} and indicative values for $g_X$ and $Y_\sigma$ corresponding to $M_X = 725\,\, \GeV$ and $M_N = 240\,\, \GeV$.

\vspace{0.6cm}
\subsection{Experimental constraints}
\label{subsec:expconst}
The three scalar fields of the present model all develop a vev and mix through the portal terms in the scalar potential \eqref{treelevelpotential}. Moreover, the corresponding mass eigenstates interact with the SM electroweak sector, as well as with the $SU(2)_X$ gauge fields and the right-handed neutrinos. The strength of these interactions is suppressed by the corresponding entries in the rotation matrix ${\cal{R}}$ \eqref{rotationmatrix}. In order to study possible signatures of the extra scalars at the LHC and future colliders, we may construct an effective Lagrangian that contains all the interactions between the scalars and the rest of the fields:
\be 
\begin{split}
\mathcal{L}_{\text{eff}}^{h_i} \, =& \, \mathcal{R}_{i1} h_i \, \left( \frac{2M^2_W}{v_h} W^+_\mu W^{-\mu} + \frac{M^2_Z}{v_h} Z_\mu Z^{\mu} - \frac{M_t}{v_h} \overline{t} t - \frac{M_b}{v_h} \overline{b} b  \right. \\
& \qquad \qquad \left. {}  - \frac{M_c}{v_h} \overline{c} c - \frac{M_\tau}{v_h} \overline{\tau} \tau + \frac{\alpha_s}{12 \pi v_h} G^a_{\mu\nu} G^{a\mu\nu} + \frac{\alpha}{\pi v_h} A_{\mu\nu} A^{\mu\nu} \right)  \\
& + \mathcal{R}_{i2} h_i \, \frac{3M_X}{v_\phi} X^a_\mu X^{a\mu} - \mathcal{R}_{i3} h_i \, \frac{M_N}{v_\sigma} \overline{N} \, N + V^h_{ijk} h_i h_j h_k , \label{ScalarEffectiveLagrangian}
\end{split}
\ee
with $V^h_{ijk}$ given by
\be 
\begin{split}
V_{ijk}^h &=   
\mathcal{R}_{i 1} \left[ \lambda_{h\phi} \mathcal{R}_{j 2} \left( v_h \mathcal{R}_{{k 2}}  + v_\phi \mathcal{R}_{{k 1}} \right) - \lambda_{h\sigma} \mathcal{R}_{{j 3}} \left( v_h \mathcal{R}_{{k 3}}  + v_\sigma \mathcal{R}_{{k 1}} \right)  \right.  \\
& \quad \quad   \left. {} +  \mathcal{R}_{{j 1}} \left( -6 \lambda_h v_h \mathcal{R}_{{k 1}}  + \lambda_{h\phi} v_\phi \mathcal{R}_{{k 2}}  - \lambda_{h\sigma} v_\sigma \mathcal{R}_{{k 3}} \right) \right]  \\
&  + \mathcal{R}_{i 2} \left[ \lambda_{h\phi} \mathcal{R}_{j 1} \left( v_h \mathcal{R}_{{k 2}}  + v_\phi \mathcal{R}_{{k 1}} \right) - \lambda_{\phi\sigma} \mathcal{R}_{{j 3}} \left( v_\phi \mathcal{R}_{{k 3}}  + v_\sigma \mathcal{R}_{{k 2}} \right)  \right.  \\
& \quad \quad  \left. {} + \mathcal{R}_{{j 2}} \left( -6 \lambda_\phi v_\phi \mathcal{R}_{{k 2}}  + \lambda_{h\phi} v_h \mathcal{R}_{{k 1}}  - \lambda_{\phi\sigma} v_\sigma \mathcal{R}_{{k 3}} \right) \right]  \\
& + \mathcal{R}_{i 3} \left[ - \lambda_{h\sigma} \mathcal{R}_{j 1} \left( v_h \mathcal{R}_{{k 3}} + v_\sigma \mathcal{R}_{{k 1}} \right) + \lambda_{\phi\sigma} \mathcal{R}_{{j 2}} \left( v_\phi \mathcal{R}_{{k 3}}  + v_\sigma \mathcal{R}_{{k 2}} \right) \right. \\
& \quad \quad  \left. {} + \mathcal{R}_{{j 3}} \left( -6 \lambda_\sigma v_\sigma \mathcal{R}_{{k 3}}  - \lambda_{h\sigma} v_h \mathcal{R}_{{k 1}} + \lambda_{\phi\sigma} v_\phi \mathcal{R}_{{k 2}} \right) \right],
\end{split}
\ee
where $i,j,k$ take the values $1,2,3$. Note that all scalar vertices containing two or more $h_2$'s are zero due to the Gildener-Weinberg conditions \eqref{MINI} and the particular parametrization of the vevs \eqref{parametrize} (see also~\cite{Farzinnia2014}). Thus, the decay rates for the decays $h_i \rightarrow h_2 h_2$ are zero at tree level. In addition, for all benchmark sets in Table \ref{table:parameters}, the decay $h_1 \rightarrow h_3 h_3$ is kinematically forbidden. Therefore, in the current framework there are not any lighter scalars that $h_1$ can decay to.

Next, let us consider the total decay widths of all scalars in relation to the corresponding SM Higgs total decay width with the same mass $ \Gamma_{h_i}^{\text{tot}} \left( M_{h} = M_{h_i} \right)$:
$$
\Gamma_{h_i}^{\text{tot}} = \mathcal{R}^2_{i1} \left[ \text{BR}^{\text{SM}}_{WW} + \text{BR}^{\text{SM}}_{ZZ} + \text{BR}^{\text{SM}}_{gg} + \text{BR}^{\text{SM}}_{\gamma\gamma} + \text{BR}^{\text{SM}}_{Z \gamma}  + \text{BR}^{\text{SM}}_{\overline{t}t} + \text{BR}^{\text{SM}}_{\overline{b}b} + \text{BR}^{\text{SM}}_{\overline{c}c} + \text{BR}^{\text{SM}}_{\overline{\tau}\tau} \right] \times \Gamma_{h}^{\text{SM}} \left( M_{h} = M_{h_i} \right) $$
\be 
\,\,\,\,\,+\,\,\,\,\, \Gamma \left( h_i \rightarrow X X \right) + \Gamma \left( h_i \rightarrow \overline{N} N \right)  + \Gamma \left( h_i \rightarrow h_j h_k \right), \label{TotalDecayRate}
\ee
where $BR^{\text{SM}}_{\chi\chi}$ are the branching ratios of the SM Higgs decays into quarks, leptons or gauge bosons. The rest of the decay rates in \eqref{TotalDecayRate} are given by
\begin{align}
\Gamma \left( h_i \rightarrow X X \right) \, &= \,  \frac{3\,M^3_{h_i}}{32\pi v^2_\phi } \, \sqrt{1 -  \frac{4 M_X^2}{M_{h_i}^2} } \, \left( 1 - 4\, \frac{M^2_X}{M^2_{h_i}} +  12\, \frac{M^4_X}{M^4_{h_i}} \right) \left| \mathcal{R}_{i 2} \right|^2 \\
\Gamma \left( h_i \rightarrow \overline{N} N \right) \, &= \,  \frac{3\,M^2_N \, M_{h_i}}{8\pi v^2_\sigma } \, \left( 1 -  \frac{4 M_N^2}{M_{h_i}^2} \right)^{3/2} \, \left| \mathcal{R}_{i 3} \right|^2 \\
\Gamma \left( h_i \rightarrow h_j h_k \right) \, &= \,  \frac{1}{16\pi} \, \frac{1}{\left( 1 + \delta_{j k} \right) M_{h_i}} \, \sqrt{1 - \frac{2(M_{h_j}^2 + M_{h_k}^2)}{M_{h_i}^2} + \frac{(M_{h_j}^2 - M_{h_k}^2)^2}{M_{h_i}^4}} \left| V^h_{i j k} \right|^2.
\end{align}
Using \Eq{TotalDecayRate}, the total decay width $\Gamma_{h_1}^{\text{tot}}$ of a SM Higgs-like scalar $h_1$ is given as
\be 
\Gamma_{h_1}^{\text{tot}} = \cos^2\alpha \cos^2\beta \, \Gamma_{h_1}^{\text{SM}} + \Gamma_{h_1}^{\text{inv}} \label{h1TotalDecayWidth},
\ee
where $\Gamma_{h_1}^{\text{SM}}$ denotes the total decay width of the SM Higgs with mass $M_{h_1} = 125.09\,\, \GeV$ and $\Gamma_{h_1}^{\text{inv}}$ is the invisible decay width of the Higgs boson to non-SM states that are kinematically allowed. Namely, only when $M_X,M_N \lesssim 62.5\,\, \GeV$ we may have
\be 
\Gamma_{h_1}^{\text{inv}} = \Gamma \left( h_1 \rightarrow X X \right) + \Gamma \left( h_1 \rightarrow \overline{N} N \right) \,.
\ee
For completeness, we present in Table \ref{table:BRs} the branching ratios of a SM Higgs with mass $M_{h_1} = 125.1\,\,\GeV$.

\vspace{0.6cm}
\begin{table}[H]
\begin{center}
\begin{tabular}{| c | c |}
\hline
Decay mode & Branching ratio \\
\hline 
$b\bar{b}$ & 0.575 \\
$W^+W^-$ & 0.216 \\
$gg$ & 0.0856 \\
$\tau^+\tau^-$ & 0.0630 \\
$c\bar{c}$ & 0.0290 \\
$Z^0Z^0$ & 0.0267 \\
$\gamma\gamma$ & $2.28 \times 10^{-3}$ \\
$\gamma Z^0$ & $1.55 \times 10^{-3}$ \\
\hline
\end{tabular}
\caption{Branching ratios for a SM Higgs boson with $M_h =125.1\,\,\GeV$, for which $\Gamma_h^{\text{SM}}=4.08\times 10^{-3}\,\,\GeV$~\cite{Heinemeyer:2013tqa}. We did not include the rest of the decay modes because their branching ratios are negligible.}
\label{table:BRs}
\end{center}
\end{table}
In order to clarify the deviation of $h_1$ from the SM Higgs, we construct the {\textit{signal strength parameter}} $\mu_{h_1}$ which can be written as
\be 
\mu_{h_1} = \frac{\sigma \left( p p \rightarrow h_1 \right)}{\sigma^{\text{SM}} \left( p p \rightarrow h \right)} \frac{\text{BR} \left( h_1 \rightarrow \chi \chi \right)}{\text{BR}^{\text{SM}} \left( h \rightarrow \chi \chi \right)} \label{SignalStrength}
\ee
where $\sigma$, $\text{BR}$ are the production cross section and branching ratio of $h_1$ and $\sigma^{\text{SM}}, \text{BR}^{\text{SM}}$ the corresponding quantities for the SM Higgs. Using \eqref{h1TotalDecayWidth} and $\sigma \left( p p \rightarrow h_1 \right) = \cos^2\alpha\cos^2\beta \,  \sigma^{\text{SM}} \left( p p \rightarrow h \right)$, the expression \eqref{SignalStrength} becomes
\be 
\mu_{h_1} = \cos^4\alpha\cos^4\beta \, \frac{\Gamma_{h_1}^{\text{SM}}}{\Gamma_{h_1}^{\text{tot}}} \label{h1SignalStrength}.
\ee
However, due to the smallness of $\mathcal{R}^2_{1 2}$ and $\mathcal{R}^2_{1 3}$, the invisible decay width $\Gamma_{h_1}^{\text{inv}}$ is highly suppressed relative to the total decay width $\Gamma_{h_1}^{\text{tot}}$ and \eqref{h1SignalStrength} simplifies to
\be
\mu_{h_1} \simeq \cos^2\alpha\cos^2\beta.
\ee
When the Higgs signal strengths from ATLAS and CMS~\cite{Cheung:2014noa,Aad:2014eva,Khachatryan:2014jba} are combined~\cite{Falkowski2015}, one obtains the constraint 
\be 
\mu_{h_1} > 0.81, \quad @\,95\% \,\, \text{C.L.},
\ee
which translates to 
\be 
\mathcal{R}_{11} = \cos\alpha\cos\beta > 0.9.
\ee
Using the benchmark values of the first set of Table \ref{table:parameters} we obtain
\be 
\mathcal{R}_{11} = 0.994,
\ee
which lies comfortably within the allowed range. Thus, the state $h_1$ behaves mostly like the SM Higgs boson and is at the moment indistinguishable from it. Run II of the LHC may be able to provide a check for the scalar sector of the model if a universal deviation for the SM Higgs couplings is established and if new scalar states are discovered.
\section{Dark matter analysis}
\label{sec:DM}
Most of the matter and energy content of the Universe still remains a mystery and has therefore been dubbed ``dark". Regarding dark matter (DM) in particular, particle physics should be able to explain its nature. In the SM alone, however, there is no viable candidate that can play the role of DM. Extending the SM, then, becomes again a necessity. On theoretical grounds, the DM particle has to be stable and produce the correct relic abundance $\Omega_{\text{DM}}$. Many direct and indirect DM detection experiments are underway or scheduled to become operational in the forthcoming years, aiming also at distinguishing between various proposed DM candidates and finally determining its properties. The befitting framework for a DM candidate should simultaneously evade limits from previous searches and be falsifiable in the near future.
\subsection{Boltzmann equation and relic density}
\label{subsec:Boltzmann}
As stated in the Introduction, the rationale behind introducing the hidden or dark $SU(2)_X$ gauge sector is twofold. First, when the new sector is spontaneously broken by means of the Coleman-Weinberg mechanism, the electroweak scale is dynamically generated through the Higgs portal. Second, since the $SU(2)_X$ gauge symmetry is completely broken by the vev $v_\phi$ of the scalar complex doublet $\Phi$, the three dark gauge bosons $X^a$ acquire equal masses $M_X=\frac{1}{2}g_X v_\phi$ and become stable due to a remnant global $SO(3)$ symmetry, thus rendering themselves potential WIMP dark matter candidates. $SU(2)_X$ vector dark matter has been studied in~\cite{Hambye2009, Zhang2010, Arina2010, Bhattacharya2012, Chiang2014, Boehm2014, Khoze2014b, Bian2015, Chen2015, Chen2015a, Gross2015, DiChiara2015a, Chen2015b} and in the context of classical scale invariance in~\cite{Hambye2013, Carone2013, Khoze2014, Pelaggi2015}.

After the end of inflation and the assumed subsequent reheating, all particles are in thermal equilibrium, while the Universe continues to expand. As the temperature continues to drop, so does the interaction rate $\Gamma_{\text{DM}}$ of the DM particles. Nevertheless, thermal equilibrium cannot be maintained and, once $\Gamma_{\text{DM}}$ becomes smaller than the expansion rate $H$ of the Universe, a DM number density ``freeze-out" occurs and dark matter particles decouple from the rest of the light degrees of freedom that remain thermalized. Thus, the DM relic abundance survives to the present epoch having the value that we observe today. Next, we calculate $\Omega_{\text{DM}}$, following Refs.~\cite{Srednicki1988a, Gondolo1991a, Kolb1990}.

\begin{figure}[H]
\centering
\includegraphics[width=13.5cm]{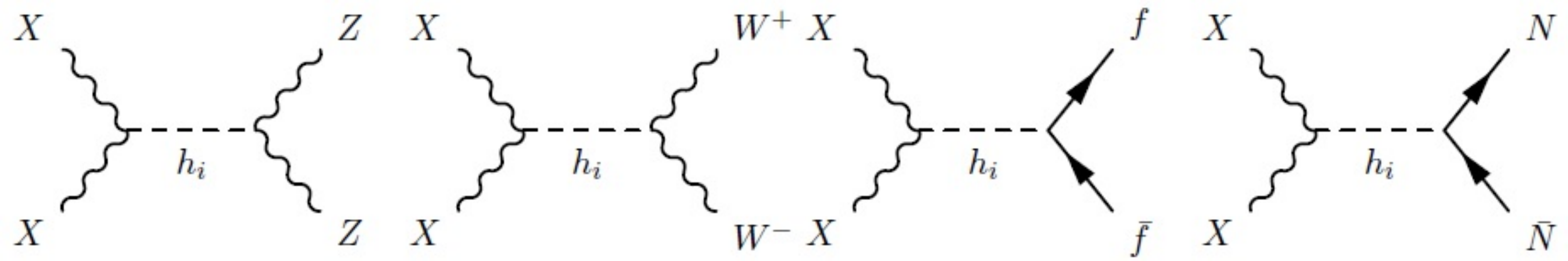}
\caption{Feynman diagrams for DM annihilation to gauge bosons and fermions.}
\label{fig:DManns1}
\end{figure}

\begin{figure}[H]
\centering
\includegraphics[width=13.5cm]{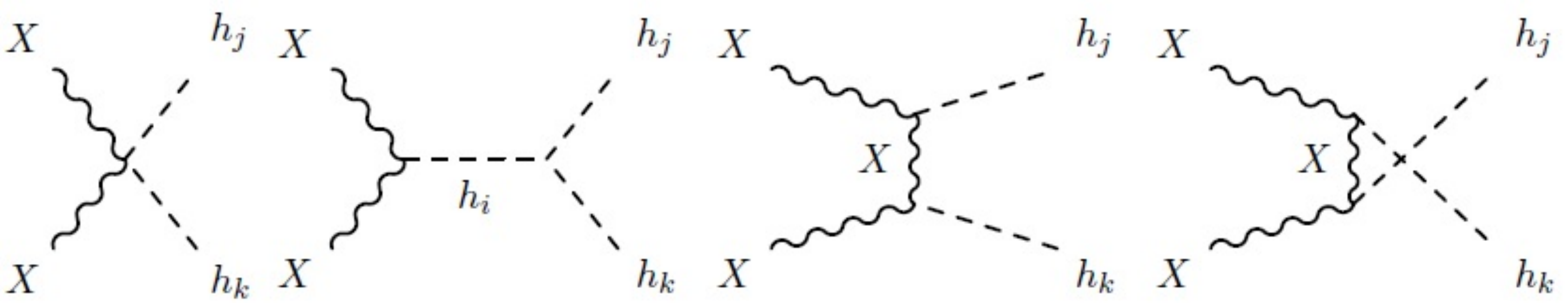}
\caption{Feynman diagrams for DM annihilation to scalars.}
\label{fig:DManns2}
\end{figure}

\begin{figure}[H]
\centering
\includegraphics[width=11.5cm]{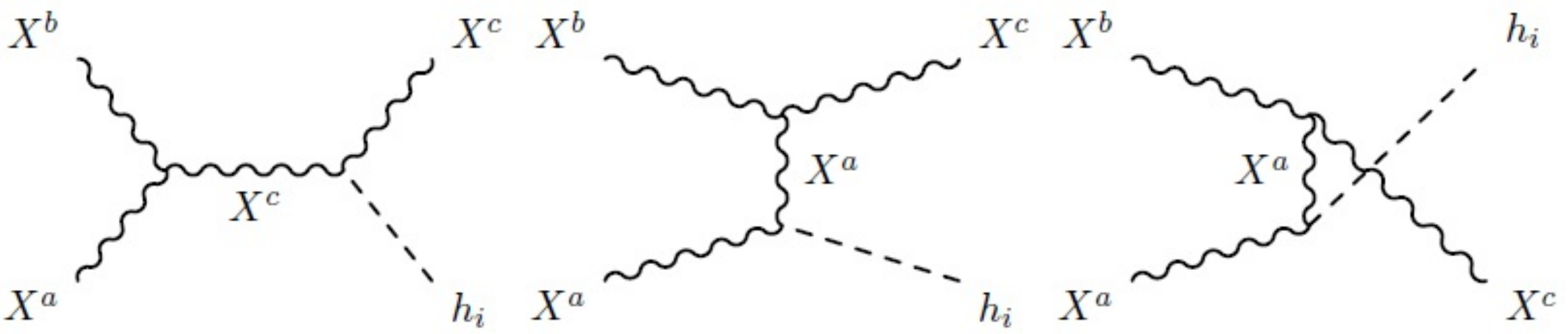}
\caption{Feynman diagrams for DM semiannihilation.}
\label{fig:DMsemi}
\end{figure}

We start with the Boltzmann equation which describes the evolution of the number density $n$ of a given particle species over time. The dark vector bosons $X^a$ can both annihilate and semiannihilate~\cite{DEramo2010,Belanger2012}, the relevant processes being listed in Figs. \ref{fig:DManns1}-\ref{fig:DMsemi}. 

\vspace{0.1cm}
\vspace{0.1cm}
The corresponding Boltzmann equation has the form~\cite{Khoze2014}
\be 
\frac{d n}{d t} + 3\,H\,n = - \, \frac{\left< \sigma v \right>_{a}}{3}  \left( n^2 - n^2_{eq} \right) - \frac{2\left< \sigma v \right>_{s}}{3} \, n \left( n - n_{eq} \right), \label{BoltzmannEquation}
\ee
where $H$ is the Hubble expansion parameter, $n_{eq}$ is the number density during equilibrium and $\left< \sigma v \right>$ is the thermally averaged cross section of the DM particles times their relative velocity, with the subscripts $a$ and $s$ denoting annihilation and semiannihilation respectively. The thermally averaged cross section times velocity is given in the nonrelativistic approximation by~\cite{Srednicki1988}
\be 
\left< \sigma v \right> \simeq \frac{1}{M^2_X} \left[ w(s) - \frac{3}{2x} \left( 2 w(s) - w'(s) \right)  \right] \Bigg\vert_{s=4M^2_X},
\ee
with the quantity $w(s)$ defined as
\be 
w(s) = \frac{1}{4} \left( 1 - \frac{\delta_{j k}}{2} \right) \beta\left( s, m_j, m_k \right) \int \frac{d( \cos\theta)}{2}\,\sum\, \lvert \mathcal{M} \left( X\,X \rightarrow \text{all} \right) \rvert^2,
\ee
where $\sum\lvert \mathcal{M} \rvert^2$ stands for the matrix element squared of all possible channels, averaging over initial polarizations and summing over final spins, $\beta\left( s, m_j, m_k \right)$ is the final-state Lorentz invariant phase space $
\beta\left( s, m_j, m_k \right) = \frac{1}{8\pi} [ 1 - ( m_j + m_k )^2/{s}  ]^{1/2} [ 1 - ( m_j - m_k)^2/{s} ]^{1/2}$ 
and $s$ denotes the usual Mandelstam variable $ 
s = ( p_1 + p_2 )^2 = 2 \left( M^2_X + E_1 E_2 - p_1 p_2 \cos\theta \right)$.
Finally, the prime stands for differentiation with respect to $s/(4M^2_X)$ and $x$ is defined as $x \equiv M_X/T$. The relevant individual cross sections necessary for the determination of the total annihilation and semiannihilation cross sections are too lengthy to write down. However, these can be found from an analogous calculation in the appendix of Ref.~\cite{Duch2015}. In Fig. \ref{fig:XS} we present these cross sections with respect to the dark matter mass $M_X$ for a fixed right-handed neutrino mass at $M_N=240\,\,\GeV$. 

\vspace{0.6cm}
\begin{figure}[H]
\centering
\includegraphics[width=11.5cm]{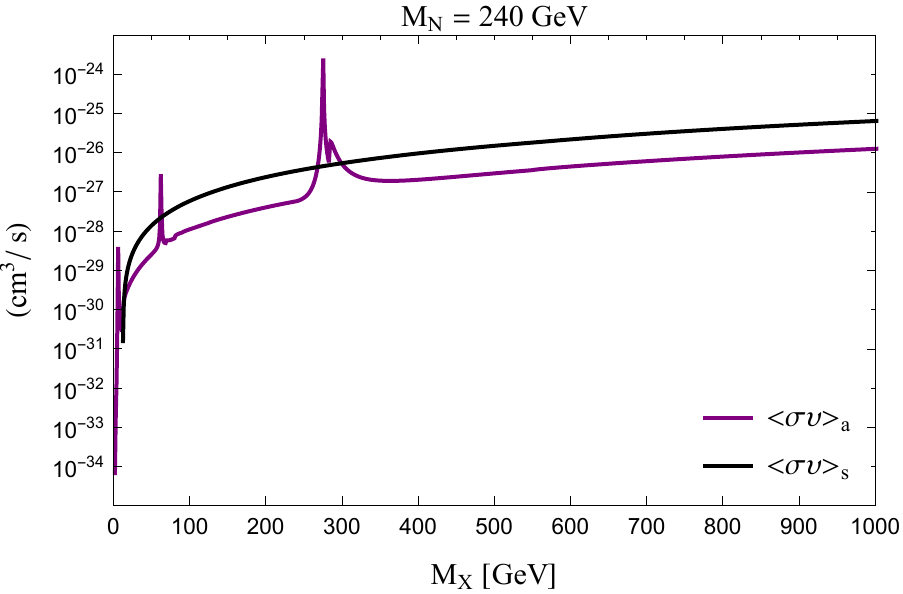}
\caption{(color online). This plot shows the thermally averaged total annihilation (purple solid line) and semiannihilation (black solid line) cross sections times relative velocity with respect to the dark matter mass $M_X$. The peaks correspond to the poles of the scalar propagators.}
\label{fig:XS}
\end{figure}

We observe that the thermally averaged semiannihilation cross section is almost an order of magnitude larger than the thermally averaged annihilation cross section. Also, we see two peaks for $\left< \sigma v \right>_{a}$ that correspond to $M_X=M_{h_1}/2$ and $M_X=M_{h_3}/2$ (for set A in Table \ref{table:parameters}) and arise due to the form of the scalar propagators at $s = 4 M^2_X$:
\be 
\Pi_{h_i} = \frac{i}{4M^2_X - M^2_{h_i} + i M_{h_i} \Gamma^{\text{tot}}_{h_i}},
\ee
with $\Gamma^{\text{tot}}_{h_i}$ given in \eqref{TotalDecayRate}. There is no peak for the darkon $h_2$ because its mass varies since it depends on $M_X$. 

Returning to the Boltzmann equation \eqref{BoltzmannEquation}, it is useful to express it in terms of the comoving volume $Y=n/\mathbf{s}$, $Y_{eq}=n_{eq}/\mathbf{s}$, where $\mathbf{s}$ is the entropy density, as
\be 
\frac{d Y}{d x} = - \frac{Z_a}{3 x^2} \left( Y^2 - Y^2_{eq} \right) - \frac{2 Z_s}{3 x^2} \left( Y^2 - Y\, Y_{eq} \right), \qquad Z_{a,s} \equiv \frac{\mathbf{s}\left( x=1 \right)}{H \left( x=1 \right)} \left< \sigma v \right>_{a,s} . \label{BoltzmannEquation2}
\ee 
The entropy density is given by $
\mathbf{s} = \frac{2 \pi^2 g_*}{45} \, \frac{M^3_X}{x^3}$ and the Hubble parameter is given by $ H = \sqrt{\frac{4 \pi^3 g_*}{45}} \, \frac{M^2_X}{M_{\text{P}}}$, in terms of the effective number of relativistic degrees of freedom $g_*$ at the time of freeze-out $\left( x = x_f \right)$. In order to solve \eqref{BoltzmannEquation2} we may consider the two extreme regions $x \ll x_f$ and $ x \gg x_f$, whereupon, defining $\Delta = Y - Y_{eq}$~\cite{Kolb1990} we obtain
\ba 
\Delta &=& - Y'_{eq} \, \frac{3 x^2}{2 \left( Z_a + Z_s \right)}, \quad \text{when} \quad x \ll x_f, \label{BoltzSol1} \\
\Delta' &=& - \frac{\Delta^2}{3 x^2} \, \left( Z_a + 2 Z_s \right), \quad \text{when} \quad x \gg x_f, \label{BoltzSol2}
\ea
where the prime now denotes $d/dx$. Moreover, if we define $\Delta(x_f) = c \, Y_{eq}(x_f)$, with $c$ being a constant of order one, we can match the solutions of \eqref{BoltzSol1} and \eqref{BoltzSol2} and obtain an expression for the freeze-out point which can be solved iteratively~\cite{Kolb1990}:
\be
x_f = \ln \left\{ 0.038 \, \frac{3 M_X M_{P}}{\sqrt{g_*(T_f) x_f}} \Big[ \, c \, (c + 2) \left< \sigma v \right>_{a} + 2 \, c \, ( c + 1 ) \left< \sigma v \right>_{s} \, \Big] \right\} . \label{fzpoint}
\ee
We find typical values between $x_f \approx 25-26$ and for the DM mass range that we consider in our numerical analysis we use $g_* = 86.25$. Also, for the constant $c$ we use $c=1/2$~\cite{Griest1991}. The present day relic abundance is obtained by integrating \eqref{BoltzSol2} from $x=x_f$ to $x=\infty$:
\be 
Y^{-1}_\infty = \int^\infty_{x_f} \frac{Z_a + 2 Z_s}{3 x^2} dx \, .
\ee
Then, using the mass density of the DM particles today, $\rho_\infty = M_X \mathbf{s}_\infty Y_\infty$ and the critical density $\rho_c = 3 H^2(\infty) M_P/(8 \pi ) = 1.054 \times 10^{-5} \, h^2 \, \mathrm{cm}^{-3}$, we finally obtain the dark matter relic density
\be 
\Omega_{X} h^2 = \frac{\rho_\infty}{\rho_c} \, h^2 = 3\times \frac{1.07\times 10^9 \, \GeV^{-1}}{\sqrt{g_*} \, M_P \, J(x_f)}, \qquad J(x_f) = \int_{x_f}^\infty dx \, \frac{\left< \sigma v \right>_{a} + 2 \left< \sigma v \right>_{s}}{x^2}, \label{RelicDensity}
\ee
where we used $\mathbf{s}_\infty = 2891.2 \,\, \mathrm{cm}^{-3}$ for the present day entropy density and $h=0.673$ for the Hubble scale factor~\cite{Olive2014a}.

\begin{figure}[H]
\centering
{\includegraphics[width=11.5cm]{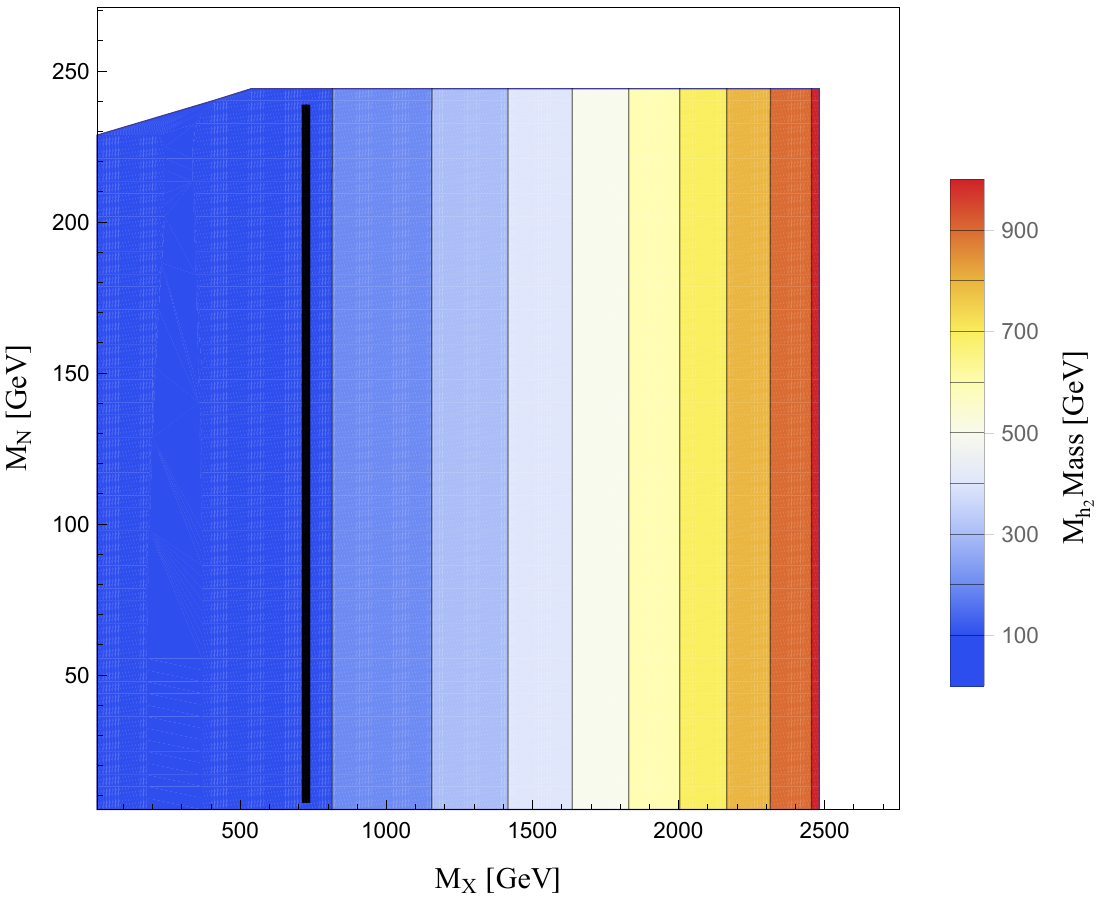}}
\caption{(color online). This plot is the same with Fig. \ref{fig:Mh2ContourPlot} but we also calculated the points where the dark gauge bosons can saturate the observed DM relic density at $3\sigma$ (black band).}
\label{fig:RelicDensity}
\end{figure}

The measured value for the DM relic density is $\Omega_{\text{DM}} h^2 \pm 1\sigma = 0.1187 \pm 0.0017$~\cite{Olive2014a}, which is a combination of the results from Planck+WP+highL+BAO. In Fig. \ref{fig:RelicDensity} we scan again the parameter space of $(g_X,Y_\sigma)$, but this time we also include the points where the DM relic density is saturated within $3\sigma$ (black region). We observe that the DM mass is constrained to be between $M_X \sim 710-740\,\,\GeV$.

\subsection{Dark matter direct detection}
\label{dirdet}
In recent years, numerous experiments have been set up aiming at directly detecting WIMP dark matter. So far these searches have not been fruitful in actually detecting dark matter. However, with each new experiment pushing the limits of sensitivity, DM detection could be just around the corner.

\begin{figure}[H]
\centering
\includegraphics[width=4.5cm]{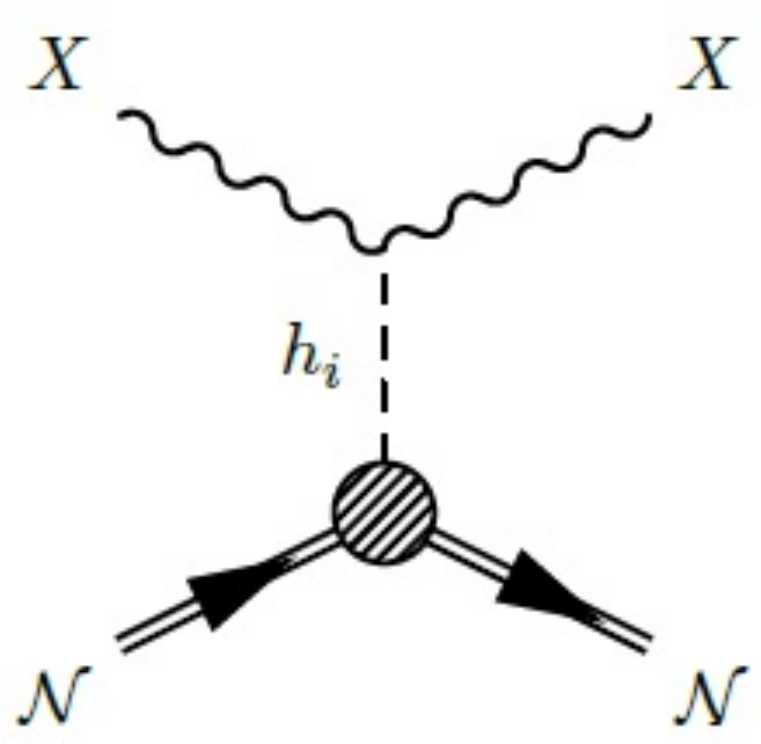}
\caption{Feynman diagram for DM-nucleon elastic scattering.}
\label{fig:DTFeynman}
\end{figure}

In the present model, the DM candidate $X$ can in principle interact with the nucleons through the $t$-channel exchange of scalar bosons $h_i$. The relevant Feynman diagram is presented in Fig. \ref{fig:DTFeynman}.

\vspace{0.6cm}

This interaction is expressed through the following effective Hamiltonian in the limit of small momentum exchange between the DM particle and the nucleon:
\be 
H_{\text{eff}} = \frac{2 M^2_X}{v_\phi} X_\mu X^\mu \left[ \sum_i \frac{\mathcal{R}_{i2} \mathcal{R}_{1i}}{M^2_{h_i}} \right] \frac{m_q}{v_h} \bar{q} q\,,
\ee
where $\mathcal{R}_{i2}$ and $\mathcal{R}_{1i}$ are the rotation matrix elements from \eqref{rotationmatrix}. The nucleonic matrix element can be parametrized as $\left< \mathcal{N} \vert \sum_q m_q \bar{q} q \vert \mathcal{N} \right> = f_{\mathcal{N}} m_{\mathcal{N}}$, where $m_{\mathcal{N}} = \left( m_p + m_n \right)/2 = 0.939\,\, \GeV$ is the average nucleon mass and $f_{\mathcal{N}}=0.303$~\cite{DiChiara2015a, Cline2013} is the nucleon form factor (see also~\cite{Junnarkar2013, Crivellin2014, Hoferichter:2015dsa}). The spin independent dark matter elastic scattering off a nucleon cross section then has the form
\be 
\sigma_{SI} = \frac{\mu_{\text{red}}^2}{\pi v_h^2 v_\phi^2} \, \left| f_{\mathcal{N}} M_X m_{\mathcal{N}}  \sum_i \frac{\mathcal{R}_{i2} \mathcal{R}_{1i}}{M^2_{h_i}} \right|^2, \label{DTXS}
\ee
where $\mu_{\text{red}} = M_X m_{\mathcal{N}}/\left( M_X + m_{\mathcal{N}} \right)$ is the DM-nucleon reduced mass, just $\mu_{red}\,\approx m_{\mathcal{N}}$ in our case.

\begin{figure}[H]
\centering
\includegraphics[width=11.5cm]{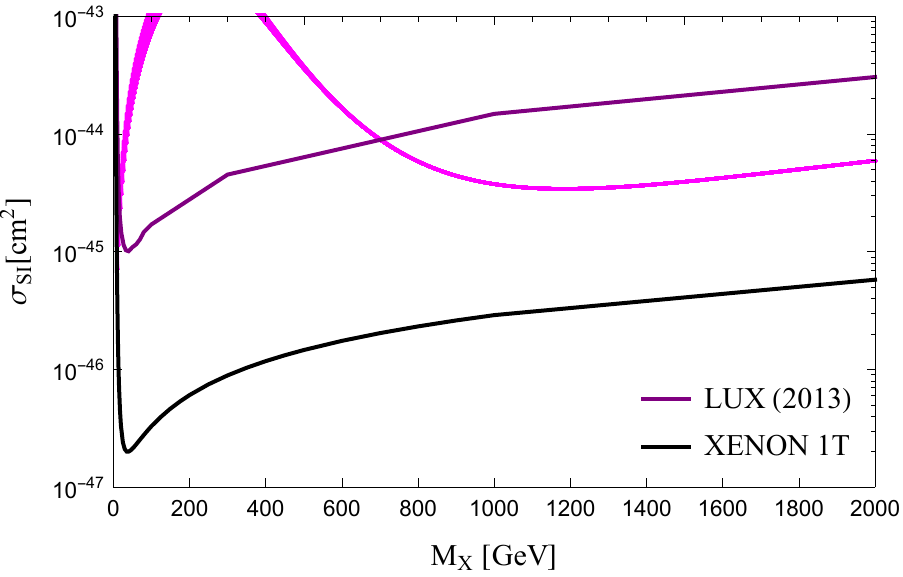}
\caption{(color online). The plot shows the DM-nucleon cross section as a function of the DM mass for varying $M_N$ masses respecting the stability and perturbativity constraints discussed in Sec. \ref{subsec:theoryconst} (magenta band). The purple solid line corresponds to the experimental limits from LUX (2013) and the black solid line corresponds to the anticipated results for XENON 1T.}
\label{fig:DTXS}
\end{figure}

Employing \eqref{DTXS}, we evaluate the spin independent cross section for various $M_X$ and $M_N$ masses and then, using the experimental results from LUX (2013)~\cite{Akerib2014a} and the projected limits from XENON 1T~\cite{Aprile2013a}, we construct the plot shown in Fig. \ref{fig:DTXS}.

\vspace{0.6cm}

We find that relatively low $M_X$ masses are excluded by LUX (2013). Nevertheless, masses above circa $700\,\,\GeV$, such as those suitable for the saturation of the measured relic density (cf. Fig. \ref{fig:RelicDensity}), are favored for detection by XENON 1T.


\section{Summary and Conclusions}
\label{sec:coclusions}
Classical scale symmetry as a framework for model building has recently received a lot of attention, mainly due to its appeal as a possible solution to the hierarchy problem through the dynamical generation of mass scales.

In this paper we considered a classically scale invariant version of the Standard Model, enlarged by a dark $SU(2)_X$ gauge group which incorporates three vector bosons and a scalar field in the fundamental representation. We also included a real singlet scalar field and Majorana neutrinos coupled to it. The dark sector was radiatively broken through the Coleman-Weinberg mechanism and a mass scale was communicated to the electroweak and neutrino sectors through the portal interactions of the dark doublet with the Higgs and singlet scalars.

We started by determining the necessary conditions for the stability of the potential and then proceeded in studying the full one-loop scalar potential, employing the Gildener-Weinberg formalism. We obtained the scalar masses through a particular parametrization of the scalar vevs, and saw that one of these masses, although zero at tree level, received large quantum corrections. Neutrinos obtained masses through the realization of a type-I low-energy seesaw mechanism.

After setting up the model, we proceeded to consider constraints to its set of free parameters through stability and perturbativity considerations. Thus, upper limits were obtained for the values of the extra gauge coupling $g_X \lesssim 2.51$ and the right-handed neutrino Yukawa coupling $Y_\sigma \lesssim 0.31$. Subsequently, the vevs and the scalar couplings were fixed by the requirement that the correct mass for the observed Higgs boson $M_{h_1} = 125.09\,\,\GeV$ is obtained. Then, the mass of one of the other scalars $M_{h_3}$ was readily computed. In order to get the last scalar mass $M_{h_2}$ (darkon), we scanned over the two-dimensional parameter space of $g_X = 2 M_X / v_\phi$ and $Y_\sigma = M_N / v_\sigma$, since the darkon's mass depends on the masses of every field present in the model (Fig. \ref{fig:Mh2ContourPlot}). In addition, we constructed an effective Lagrangian describing the interactions of the mixed scalars with the rest of the fields, where cubic terms of the form $h_i - h_2 - h_2$ were not present due to the Gildener-Weinberg minimization conditions \eqref{MINI}. Since for the benchmark sets of values in Table \ref{table:parameters} the mass of $h_3$ turned out to be larger than the $h_1$ mass, we concluded that in the current setup the Higgs boson does not decay to any lighter states. Afterwards, we calculated the signal strength parameter corresponding to $h_1$ and found that the common suppression factor $\mathcal{R}_{ 1 1 }$ of the Higgs' couplings with the SM fields is in agreement with bounds set by LHC.

Finally, we identified the extra gauge bosons as WIMP dark matter candidates. We considered the Boltzmann equation and solved it semianalytically in the nonrelativistic approximation after calculating the total thermally averaged annihilation and semiannihilation cross sections (Fig. \ref{fig:XS}). In this way we obtained the dark matter relic density, and by matching it to the observed value we constrained the dark matter mass to be in the range $M_X \sim 710 - 740 \,\, \GeV$ (Fig. \ref{fig:RelicDensity}). Then, considering the dark matter elastic scattering off a nucleon we computed the spin-independent scattering cross section and compared it with existing and projected limits from direct detection experiments. We found that dark matter masses above $\sim 700\,\, \GeV$ evade limits set by LUX (2013) but can nevertheless be tested in the next years by XENON 1T.

In conclusion, the classically scale invariant model that we considered is a perfectly viable extension of the Standard Model, able to dynamically generate the dark matter, neutrino and electroweak scales through the multi-Higgs portal while stabilizing the vacuum. It predicts new scalar states that future collider searches may be able to discover and predicts vector dark matter with a definite mass range that can be probed by direct detection experiments in the years to come.
\section*{Acknowledgements}
This research has been cofinanced by the European Union (European Social Fund - ESF)
and Greek national funds through the Operational Program Education and Lifelong
Learning of the National Strategic Reference Framework (NSRF) - Research Funding
Program: \textit{ARISTEIA - Investing in the society of knowledge through the European Social Fund.} K.T. would also like to thank I. Antoniadis and K. Papadodimas for discussions and hospitality at the CERN Theory Division. A.K. would like to thank Gunnar Ro and Dimitrios Karamitros for useful discussions and correspondence. 

\bibliography{C:/Users/ALEX/Dropbox/1stPaper/References}{}
\bibliographystyle{utphys}
\end{document}